\documentclass[prd,preprint,superscriptaddress,preprintnumbers,eqsecnum,showpacs,nofootinbib,nobibnotes]{revtex4}
\usepackage{amsfonts,amsmath,amssymb,bm,natbib}
\usepackage{graphicx} 
\usepackage{pstricks}

\newcommand{\be}{\begin{equation}}
\newcommand{\bea}{\begin{eqnarray}}
\newcommand{\ee}{\end{equation}}
\newcommand{\eea}{\end{eqnarray}}
\def\s#1{{\scriptscriptstyle #1}}

\def\1eq#1{Eq.~(\ref{#1})}

\def\2eqs#1#2{Eqs.~(\ref{#1}) and~(\ref{#2})}
\def\3eqs#1#2#3{Eqs.~(\ref{#1}),~(\ref{#2}) and~(\ref{#3})}

\def\chic#1{{\scriptscriptstyle #1}}

\def\fig#1{Fig.~\ref{#1}}

\def\ie{{\it i.e.}, }

\def\Deltam{\Delta_{m}}

\def\Vqqq{V}

\def\Jm{J_m}

\def\s#1{{\scriptscriptstyle #1}}

\def\n#1{({\it #1}\,)}

\def\gA{g^2 C_A}

\begin{document}

\title{Gluon mass generation in the massless bound-state formalism}

\author{D. Iba\~nez}

\author{J. Papavassiliou}
\affiliation{\mbox{Department of Theoretical Physics and IFIC, 
University of Valencia and CSIC},
E-46100, Valencia, Spain}

\begin{abstract}

We present a detailed, all-order study of gluon mass generation within
the  massless  bound-state formalism,  which  constitutes the  general
framework for the systematic implementation of the Schwinger mechanism
in non-Abelian gauge theories.   The main ingredient of this formalism
is the dynamical formation  of bound-states with vanishing mass, which
give  rise  to effective  vertices  containing  massless poles;  these
latter vertices,  in turn, trigger the Schwinger  mechanism, and allow
for the  gauge-invariant generation of an effective  gluon mass.  This
particular approach has the conceptual advantage of relating the gluon
mass  directly to  quantities that  are intrinsic  to  the bound-state
formation  itself,  such  as  the  ``transition  amplitude''  and  the
corresponding  ``bound-state   wave-function''.   As  a   result,  the
dynamical  evolution of  the gluon  mass  is largely  determined by  a
Bethe-Salpeter  equation that  controls the  dynamics of  the relevant
wave-function, rather than the  Schwinger-Dyson equation of the gluon
propagator,  as  happens  in  the  standard  treatment.   The  precise
structure and  field-theoretic properties of  the transition amplitude
are scrutinized in  a variety of independent ways.  In particular, a 
parallel study within  the linear  covariant (Landau)  gauge  and the
background-field method  reveals that a powerful identity, 
known to be valid at the level of conventional Green's functions,  
relates also the background and quantum transition  amplitudes.
Despite  the differences in the  ingredients and terminology employed, the
massless  bound-state  formalism   is  absolutely  equivalent  to  the
standard approach based on  Schwinger-Dyson equations.  In fact, a set
of powerful relations allow  one to demonstrate the exact coincidence
of  the integral  equations governing  the momentum  evolution  of the
gluon mass in both frameworks.

\end{abstract}


\maketitle

\section{Introduction}

The dynamical generation of an effective (momentum-dependent) 
gluon mass in pure Yang-Mills theories~\cite{Cornwall:1981zr,Bernard:1982my,Donoghue:1983fy} 
(and eventually in QCD) has attracted considerable attention in recent years, 
because it furnishes a cogent theoretical explanation for a large number of 
important lattice findings~\cite{Cucchieri:2007md,Cucchieri:2007rg,Cucchieri:2009zt,Bogolubsky:2007ud,Bowman:2007du,Bogolubsky:2009dc,Oliveira:2009eh,Iritani:2009mp}. 
This inherently nonperturbative phenomenon is usually studied  
within the formal machinery of the Schwinger-Dyson equations (SDEs), 
supplemented by a set of fundamental guiding principles, which enable 
the emergence of ``massive'' solutions, while preserving intact the 
gauge invariance of the theory~\cite{Aguilar:2008xm,Aguilar:2011ux}.

The main theoretical concept underlying the existence of such solutions 
is the Schwinger mechanism~\cite{Schwinger:1962tn,Schwinger:1962tp}.
The basic observation is that if the ``vacuum polarization''   
acquires a pole at zero momentum transfer, then the 
 vector meson becomes massive, even if the gauge symmetry 
forbids a mass at the level of the fundamental Lagrangian, as happens in the case of gauge theories.
In the absence of elementary scalar fields,  
the origin of the aforementioned poles must be purely nonperturbative: for sufficiently strong binding, 
the mass of certain (colored) bound states 
may be reduced to zero~\cite{Jackiw:1973tr,Jackiw:1973ha,Cornwall:1973ts,Eichten:1974et,Poggio:1974qs}.  
In addition to  triggering the Schwinger mechanism, these bound-state poles 
act as composite, longitudinally coupled Nambu-Goldstone bosons, enforcing the 
gauge invariance of the theory in the presence of gauge boson masses.  
Every such Goldstone-like scalar, ``absorbed'' by
a gluon in order to acquire a mass,  
is expected to actually cancel out of the $S$-matrix   
against other massless poles or due to current conservation.
 
The concrete realization of the above general  
scenario in Yang-Mills theories proceeds by appealing to  
the existence of a special type of nonperturbative vertices, to be generically denoted by $V$ 
(also referred to as ``pole vertices''), which essentially 
transmit the effects of the composite excitations to the dynamical equations governing the 
various off-shell Green's functions.
Specifically, the inclusion of these vertices  
\n{i} enables the SDE of the gluon propagator to admit massive solutions, and  
\n{ii}  guarantees that the Ward identities (WIs) and the Slavnov-Taylor identities 
(STIs) of the theory maintain exactly the same form before and after gluon mass generation; 
a particular consequence of this important property is the exact transversality of 
the resulting (massive) gluon self-energy. 

As has been demonstrated in a recent work~\cite{Binosi:2012sj}, 
the full dynamical equation of the gluon mass in the Landau gauge
may be derived from the SDE of the gluon propagator by 
postulating the existence of the pole vertices, and employing  the WIs and STIs they satisfy, 
together with their totally longitudinal nature 
(for related studies in the Coulomb gauge, 
see~\cite{Szczepaniak:2001rg,Szczepaniak:2003ve,Epple:2007ut,Szczepaniak:2010fe}). 
The resulting integral equation makes no reference 
to the closed form of these vertices, nor 
to the actual dynamical mechanism responsible for their existence.        
However, the details of the actual dynamical formation of the bound-state poles, 
and subsequently of the pole vertices, 
as well as their exact closed form,  
transcend the concrete purpose of obtaining a particular version of the gluon mass equation, and 
are of  paramount importance
for the self-consistency of the entire     
gluon mass generation scenario. In fact, it would be highly desirable to 
establish a precise quantitative connection between the fundamental ingredients 
composing these vertices and the gluon mass itself.

The purpose of the present work is to 
dissect the pole vertices and scrutinize the field-theoretic properties of their 
constituents, within the context of the ``massless bound-state formalism'',  
first introduced in some early seminal contributions 
to this subject~\cite{Jackiw:1973tr,Jackiw:1973ha,Cornwall:1973ts,Eichten:1974et,Poggio:1974qs}, 
and further developed in~\cite{Aguilar:2011xe}. The final outcome of this analysis is an alternative, but completely equivalent,
description of the dynamical gluon mass in terms of quantities appearing naturally in the physics of bound states, 
such as the ``transition amplitude'' and the  ``bound-state wave function''. This new description 
makes manifest some of the salient  physical properties of the gluon mass 
({\it e.g.}, positive-definiteness), 
and provides, in addition, a decisive confirmation of the self-consistency of the concepts and methodology employed. 

Let us next present the general outline of the article, introducing some of the basic concepts, 
and commenting on the logical connections and delicate interplay between the various sections.

The starting point of our considerations is a brief review of 
the importance of the pole vertices for obtaining 
 massive solutions out of the SDE for the gluon propagator, 
in a gauge invariant way, \ie preserving the form of the fundamental STIs of the theory (Section \ref{massde}).

Within the massless bound-state formalism, the pole vertices 
are composed of three fundamental ingredients. 
\n{i} The nonperturbative transition amplitude, to be denoted by $I(q^2)$,   
which connects a single gluon to the massless excitation.  
\n{ii} The scalar massless excitation, whose propagator furnishes the pole ${i}/{q^2}$, and 
\n{iii} a set of 
``proper vertex functions''~\cite{Jackiw:1973ha} (or ``bound-state wave functions''),
to be generically denoted by $B$, (with appropriate Lorentz and color indices), which 
connect the massless excitation to a number of gluons and/or ghosts.  
The quantity $I(q^2)$ is 
universal, in the sense that it appears in all possible pole vertices. Furthermore,  
it admits its own diagrammatic representation, which, in turn, involves the functions $B$.
As a result, the dependence of the pole vertices on the $B$s is quadratic (Section~\ref{mbsf}).

When inserted into the SDE for the gluon propagator, and all diagrams are kept (no truncation), 
the special structure of the pole vertices allows one to  obtain a very concise relation 
between the gluon mass and the square of the transition amplitude, given in \1eq{QQmassformula}. 
This relation demonstrates that, 
unless $I(q^2)$ vanishes identically, the gluon mass obtained is positive-definite (Section~\ref{gmta}).

If the above construction is repeated within the combined  framework of  
the pinch technique (PT)~\cite{Cornwall:1981zr,Cornwall:1989gv,Pilaftsis:1996fh,Binosi:2002ft,Binosi:2003rr,Binosi:2009qm}  
and the background field method (BFM)~\cite{Abbott:1980hw}  
(known as the ``PT-BFM scheme''~\cite{Aguilar:2006gr,Binosi:2007pi,Binosi:2008qk}), 
the relevant quantity to consider is the transition amplitude between a {\it background gluon} and the  
massless excitation, to be denoted by $\widetilde{I}(q^2)$ (all other ingredients remain identical). 
The mass of the propagator connecting a background ($B$) and a quantum gluon ($Q$) 
can then be expressed as the product of $I(q^2)$ and $\widetilde{I}(q^2)$;  
an equivalent relation may also be obtained from the background-background propagator ($BB$). 
The additional fact that the PT-BFM gluon propagators ($QB$ and $BB$)
are related to the conventional one ($QQ$) by a set of   
powerful identities (known as Background-Quantum identities (BQIs)~\cite{Grassi:1999tp,Binosi:2002ez}), 
allows one finally to relate $I(q^2)$ and $\widetilde{I}(q^2)$ as shown in \1eq{BQItransition} (Section~\ref{bqi}).

Evidently, \1eq{BQItransition} has emerged as a self-consistency requirement between 
two different formulations of the SDEs (conventional and PT-BFM), 
which, in their untruncated version, must furnish the same physics.
It would be very important, however, to establish the validity of \1eq{BQItransition} 
in a more direct, explicit way, by operating at the level of the 
pole vertices themselves, where  $I(q^2)$ and $\widetilde{I}(q^2)$ make their primary appearance.
To that end, we will carry out the explicit construction of the three-gluon pole vertex,
both in the conventional and the BFM formalism,  
using as a sole input  the WI and/or STIs they satisfy, and their 
totally longitudinal nature (subsection \ref{expcon}).  
The comparison of the two, after judicious identification of the parts that 
contribute to the gluon mass equation, reproduces precisely \1eq{BQItransition} 
(subsection \ref{upart}).

It turns out that an even more fundamental derivation of \1eq{BQItransition} may be devised, 
which takes one back to the underpinnings of the PT-BFM connection: 
the BQIs, which are formally obtained within the Batalin-Vilkovisky (BV)
formalism~\cite{Batalin:1977pb,Batalin:1981jr}, can be alternatively derived through the 
diagrammatic rearrangements implemented by the PT (Section \ref{bqiexp}). 
In the case of the SDE series containing regular (fully dressed) vertices, 
such a diagrammatic derivation amounts finally to the demonstration of the nonperturbative  
PT-BFM equivalence~\cite{Binosi:2008qk}. From the operational point of view, the 
standard PT construction boils down to the 
judicious exploitation of the rearrangements produced when certain longitudinal (pinching)
momenta trigger the STIs satisfied by the aforementioned vertices;  
the required STIs are known from the formal machinery of the BV, or alternative formalisms.
A priori, the implementation of this procedure at the 
level of the diagrammatic representation of the $I(q^2)$ is thwarted by the fact that 
the pinching momenta will act on the $B$ vertices (introduced above), whose STIs, however,
are not formally known. Nonetheless, the explicit construction of the pole vertex, 
and the subsequent line of reasoning, furnish precisely the missing STI, and make the 
PT-driven diagrammatic construction possible (subsection \ref{dem}). 
An interesting by-product of this construction is the  
derivation of an integral constraint between the $B$-vertex functions containing ghost legs. 
When combined with the recent lattice findings on the 
infrared behavior of the ghost propagator, this constraint strongly suggests the individual 
vanishing of all such ghost vertex functions (subsection \ref{ghostcon}).

The diagrammatic evaluation of the transition amplitude furnishes \1eq{completefulltransition}, 
which expresses this fundamental ingredient of the bound-state formalism in terms of a double 
integral containing solely the vertex $B_{\mu\nu}$. Even though this vertex function is not 
known for general momenta, the powerful relation of \1eq{relB1mass}, 
first derived in~\cite{Aguilar:2011xe}, relates its $g_{\mu\nu}$ 
form factor to the gluon mass. This relation, in turn, allows one to recover 
the mass equation derived in~\cite{Binosi:2012sj},
following a completely different methodology and formalism. The resulting exact coincidence 
reveals an impressive complementarity between two formally distinct methods (subsection~\ref{self}).

The equivalence of the two formalisms  
established in the previous subsection is exact only when no approximations are 
employed in either one. However, 
in practice approximations must be carried out, and as a result, in general, 
due to the difference of the intermediate steps, 
a modified momentum dependence for the gluon masses will be obtained. 
In subsection \ref{VS} we outline the general  
computational procedure that must be followed within the massless bound-state formalism
in order to determine the dynamical gluon mass. 
The nature of this procedure is {\it a-priori} dissimilar to 
that employed in~\cite{Binosi:2012sj}; thus, whereas an SDE equation is 
solved in the latter case, now the main dynamical ingredient is the 
Bethe-Salpeter equation (BSE) that controls the evolution of the relevant bound-state wave function.
In the qualitative discussion presented here,  
we focus on the difference in the momentum-dependence of the gluon mass 
that one observes at the lowest level of approximation within both formalisms. 

Finally, our conclusions and discussion are presented in Section~\ref{concl} .

\section{\label{massde}Getting massive solutions from the gluon SDE}

In this section we review the general principles that allow the generation of massive solutions 
out of the gluon SDE, and study in detail the general structure of the special pole 
vertices that trigger this effect (for different approaches, see, {\it e.g.},~\cite{Aguilar:2002tc,
Aguilar:2004sw,Dudal:2008sp,Boucaud:2008ky,Fischer:2008uz,RodriguezQuintero:2010ss,
Bicudo:2010wi,Oliveira:2010xc,Kondo:2011ab,Qin:2011dd,Gonzalez:2011zc,Pennington:2011xs,Bashir:2011dp,Bashir:2012fs,Kondo:2012ri,Strauss:2012dg}).

\subsection{General principles}

\begin{figure}[t!]
\center{\includegraphics[scale=0.6]{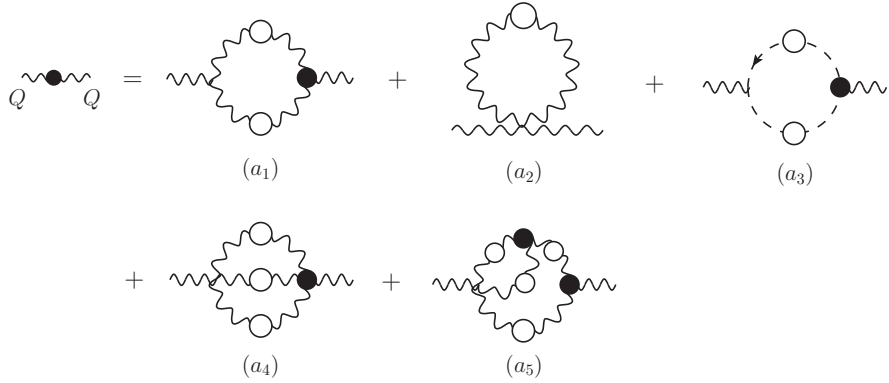}}
\caption{SDE satisfied by the conventional 
gluon self-energy, with two quantum gluons ($Q$) entering. White (black) circles denote fully-dressed propagators (vertices).}
\label{QQ-SDE}
\end{figure}

The full gluon propagator 
$\Delta^{ab}_{\mu\nu}(q)=\delta^{ab}\Delta_{\mu\nu}(q)$ in the Landau gauge is defined as
\be
\Delta_{\mu\nu}(q)=- i P_{\mu\nu}(q)\Delta(q^2) \,,
\label{prop}
\ee 
where
\be
P_{\mu\nu}(q)=g_{\mu\nu}- \frac{q_\mu q_\nu}{q^2} \,,
\ee
is the usual transverse projector, 
and the scalar cofactor $\Delta(q^2)$  
is related to the (all-order) gluon self-energy $\Pi_{\mu\nu}(q)=P_{\mu\nu}(q)\Pi(q^2)$  through
\be
\Delta^{-1}({q^2})=q^2+i\Pi(q^2).
\label{defPi}
\ee
Alternatively, one may introduce the {\it inverse} of the gluon dressing function, $J(q^2)$, defined as~\cite{Ball:1980ax}
\be
\Delta^{-1}({q^2})=q^2 J(q^2) \,.
\label{defJ}
\ee
Let us 
consider the SDE for the gluon propagator in the conventional formulation of Yang-Mills (\ie linear $R_{\xi}$ gauges) 
\begin{equation}\label{QQSDE}
\Delta^{-1}(q^2) P_{\mu\nu}(q) = q^2P_{\mu\nu}(q)+i\Pi_{\mu\nu}(q) ,
\end{equation}
with the self-energy given by 
\begin{equation}
\Pi_{\mu\nu}(q)=\sum_{i=1}^5(a_i)_{\mu\nu}\,,
\label{selfen}
\end{equation}
where the diagrams $(a_i)$ are shown in Fig.~\ref{QQ-SDE}.
Note that the full (untruncated) self-energy is transverse, namely 
\begin{equation}
q^{\mu} \Pi_{\mu\nu}(q)=\sum_{i=1}^5 q^{\mu} (a_i)_{\mu\nu} = 0 \,.
\label{selfentran}
\end{equation}

It turns out that, with the Schwinger mechanism turned off (\ie in the absence of massless poles in the vertices) 
the SDE leads to the conclusion that $\Delta^{-1}({0}) =0$, 
namely the absence of massive solutions. 

In order to obtain massive solutions out of the above SDE,  
and preserve, at the same time, the gauge invariance intact, one must carry 
out the crucial substitution 
\begin{equation}
\label{modifiedvertex}
\Gamma \longmapsto \Gamma' = \Gamma\!_m + V,
\end{equation} 
to all fully-dressed interaction vertices appearing in \1eq{QQSDE}. 
The main characteristic of the vertices $V$, 
which sharply differentiates them from ordinary vertex contributions, 
is that they contain massless poles, originating from the 
contributions of bound-state excitations. 
Such dynamically generated poles are to be 
clearly distinguished from poles related to  
ordinary massless propagators, associated with elementary fields in the original Lagrangian.
In addition, they are completely {\it longitudinally} coupled, 
\ie they satisfy 
 conditions of the type (for the case of the three-gluon vertex) 
\be
P^{\alpha'\alpha}(q) P^{\mu'\mu}(r) P^{\nu'\nu}(p) \Vqqq_{\alpha'\mu'\nu'}(q,r,p)  = 0.
\label{totlon}
\ee   
As for the vertices $\Gamma\!_m$, 
they are given by the same graphs as the $\Gamma$ before, but 
with gluon propagators replaced by massive ones [see \1eq{massive}], 
implementing simultaneously \1eq{modifiedvertex}.

The new (massive) self-energy is then given by    
\begin{equation}
\Pi_{\mu\nu}(q)= \sum_{i=1}^5(a'_i)_{\mu\nu},
\label{selfenmas}
\end{equation}
where the ``prime'' indicates that the various fully-dressed 
vertices appearing inside the corresponding diagrams of the gluon self-energy 
have been replaced by their primed counterparts, as in \1eq{modifiedvertex}. 
It is important to emphasize that, since the above replacement maintains the STIs of the theory unaltered,   
the transversality of the massive self-energy persists, {\it i.e.}, 
\begin{equation}
\sum_{i=1}^5 q^{\mu} (a'_i)_{\mu\nu} = 0 \,.
\label{selfenmastran}
\end{equation}

The appearance of massive solutions amounts effectively 
to the change   
(in Minkowski space)
\be
\Delta^{-1}(q^2) = q^2 J(q^2) \longmapsto \Deltam^{-1}(q^2)=q^2 \Jm(q^2)-m^2(q^2),
\label{massive}
\ee
with $m^2(0) \neq 0$ (of course, in Euclidean space, one must find $\Deltam^{-1}(0) >0$). 
The subscript ``m'' in $\Jm$
indicates that effectively one has now a mass inside the various expressions: for example, 
whereas perturbatively $J(q^2) \sim \ln q^2$,
after dynamical gluon mass generation has taken place, one has $J_m(q^2) \sim \ln(q^2+m^2)$.

The actual evaluation 
of the relevant diagrams 
may be carried out by appealing to the basic  
global features of the $V$ vertices, as deduced from the
STIs and their complete longitudinality~\cite{Binosi:2012sj}.
The final upshot is that 
the SDE may be schematically cast into the form (Minkowski space)
\be
q^{2} J_m(q^2) - m^{2}(q^2) =  
q^{2} \left[1+{\cal K}_1 (q^{2},m^2,\Delta_m)\right] + {\cal K}_2 (q^{2},m^2,\Delta_m),
\ee
such that $q^{2} {\cal K}_1 (q^{2},m^2,\Delta_m) \to 0$, as $q^{2}\to 0$, 
whereas ${\cal K}_2 (q^{2},m^2,\Delta_m)\neq 0$ in the same limit, 
precisely because it includes the term $1/q^2$ contained inside $V_{\alpha\mu\nu}(q,r,p)$.
This form, in turn, gives rise to 
two coupled integral equations, 
an inhomogeneous equation for $J_m(q^2)$, and a homogeneous one for $m^{2}(q^2)$  
(the latter is usually referred to as the ``mass equation''),
of the generic type
\bea
J_m(q^2) &=& 1+ \int_{k} {\cal K}_1 (q^{2},m^2,\Delta_m),
\label{Jeq}\\
m^{2}(q^2) &=&  - \int_{k} {\cal K}_2 (q^{2},m^2,\Delta_m).
\label{meq}
\eea
Physically meaningful (\ie positive definite and monotonically decreasing) 
solutions to an approximate version of the full mass equation have been 
recently presented in~\cite{Binosi:2012sj}.

\subsection{\label{strV}Structure of the vertices $V$}

We will next consider the decomposition of the vertices $V$ that emerges in a natural way, 
if one employs as a criterion the effect that the various components of $V$ 
may have on the gluon SDE. 
For concreteness we will focus on the case $V_{\alpha\mu\nu}$; however, all basic arguments 
may be straightforwardly extended to any vertex of the type $V$.  

Since the main function of the vertex $V_{\alpha\mu\nu}$ is to generate a mass term when inserted 
into the graph ($a_1$)  of Fig.~\ref{QQ-SDE}, it must contain components that 
do not vanish as $q\rightarrow 0$. To study this point in more detail, 
we first separate $V$ into two distinct parts, 
namely 
\begin{equation}\label{Vgeneric}
V_{\alpha\mu\nu}(q,r,p) = {U}_{\alpha\mu\nu}(q,r,p) + {R}_{\alpha\mu\nu}(q,r,p)\,,
\end{equation}
defined as follows.
${U}$ is the part of $V$ that has its Lorentz  index $\alpha$  
saturated by the momentum $q$; thus, it contains necessarily 
the explicit $q$-channel massless excitation,  namely the $1/q^2$ poles. 
It assumes the general form 
\be
{U}_{\alpha\mu\nu}(q,r,p) = \frac{q_{\alpha}}{q^2} C_{\mu\nu}(q,r,p), 
\label{Unoncal}
\ee
where, due to Bose symmetry under the exchange $r\leftrightarrow p$, $\mu \leftrightarrow \nu$, 
$C_{\mu\nu}(q,r,p)$ must satisfy \mbox{$C_{\nu\mu}(q,p,r) = - C_{\mu\nu}(q,r,p)$}; 
as a result, $C_{\mu\nu}(0,-p,p)=0$.

The term ${R}$ contains everything else; in particular, 
the massless excitations in the other two kinematic channels, 
namely $1/r^2$ and $1/p^2$ (but not $1/q^2$) are assigned to ${R}$.
Thus, for example, terms of the form $q_{\alpha} g_{\mu\nu}$ and 
$q_{\alpha} r_{\mu} p_{\nu}$ are assigned to ${U}_{\alpha\mu\nu}$, while terms of the type 
$p_{\nu} g_{\alpha\mu}$ and $r_{\mu} g_{\alpha\nu}$ belong to ${R}$. 
Evidently, $P^{\mu'\mu}(r) P^{\nu'\nu}(p){R}_{\alpha \mu'\nu'}(q,r,p) = 0$.
In addition, again due to Bose symmetry, we have that ${R}_{\alpha \mu\nu}(0,-p,p) = 0$. 

Consider now the individual effect that ${U}$ and ${R}$ have when inserted into the gluon SDE, 
specifically the graph ($a_1$)  of Fig.~\ref{QQ-SDE}. 
Of course, it is expected, on general grounds, that ${R}$  should not  
generate mass terms, because it does not contain poles of the type $1/q^2$;  
this is indeed what happens. 
If we work in the Landau gauge, any contribution from $R$ vanishes for a simple kinematic reason, 
namely the above transversality condition satisfied by ${R}$. 
Away from the Landau gauge, ${R}$ does not contribute to the 
mass equation \1eq{meq} either, because ${R}_{\alpha \mu\nu}(0,-p,p) = 0$  
and there is no $1/q^2$ term that could compensate this. 
Thus, ${R}$ seems to contributes, in a natural way, to the equation for $J_m(q^2)$.

In fact, it is relatively straightforward to establish  
that, in the limit $q\rightarrow 0$, the SDE 
contribution generated by ${R}$ vanishes as ${\cal O}(q^{2})$ (or faster).
Indeed, let us set $p=k$, the virtual integration momentum in the SDE diagram; 
given that ${R}_{\alpha \mu\nu}(0,-k,k) = 0$, a  Taylor expansion of ${R}$ around $q^2=0$ gives 
(suppressing indices)
\be
 R(q,-q-k,k) = 2(qk) R'(k^2) + {\cal O}(q^2)\,.
\ee
Now, the first term is odd in $k$, and therefore the corresponding integral vanishes. As a result,  
the term $2(qk) R'(k^2)$ must be multiplied by another term proportional to $(qk)$, coming from the rest of the 
terms (propagators, vertices, etc) appearing in the integrand. Thus, the resulting contribution is 
of order ${\cal O}(q^{2})$ (or higher), as announced. The importance of this property is related to the fact that, 
in order to arrive at \1eq{Jeq}, one must pull out of the corresponding integral equation a factor of $q^2$;  
had the integrand vanished slower than ${\cal O}(q^{2})$, one would get a divergent contribution for $J_m(0)$, which would be 
physically unacceptable, given that $J_m(0)$ is inversely proportional to the infrared finite QCD effective charge
[see, {\it e.g.},~\cite{Aguilar:2009nf}, and references therein].

Therefore, the only term that can contribute to the gluon mass equation is ${U}_{\alpha\mu\nu}(q,r,p)$;  
the precise contribution will depend, of course, on the exact form and behaviour 
of the cofactor $C_{\mu\nu}(q,r,p)$, as $q\rightarrow 0$. 
It is clear for instance, that if $C_{\mu\nu}(q,r,p)$ contains terms that  
behave as ${\cal O}(q^{1+c})$,  with $c>0$, as $q\rightarrow 0$, then these terms 
could not possibly trigger the Schwinger mechanism, because the effect of the pole would be counteracted 
by the positive powers of $q$.
On the other hand, terms that vanish as ${\cal O}(q^{1-c})$, 
would give rise to divergent results; 
however, this latter possibility does not occur, again due to the 
Bose symmetry of $V$ with respect to $p \leftrightarrow r$. 
Thus, the only terms of $C_{\mu\nu}$ relevant for 
gluon mass generation are those that vanish as ${\cal O}(q)$ $(c=0)$.

These observations motivate the separation of ${U}_{\alpha\mu\nu}(q,r,p)$ into two parts, 
one that behaves as a constant, ${\cal O}(q^{0})$, thus contributing to the mass equation  
(to be denoted by ${\cal U}_{\alpha\mu\nu}$), 
and one that vanishes as ${\cal O}(q^{c})$, $c>0$, and can be naturally reassigned to ${R}$
(to be denoted by ${\cal U^{\,\prime}_{\alpha\mu\nu}}$).
In fact, due to the same reason explained above, namely, the absence of divergent contributions at the level of \1eq{Jeq}, 
we must have that $c \geq 1$; actually, the explicit construction presented in Section~\ref{excon} [{\it e.g.} \1eq{quantumHT} and ensuing discussion], 
reveals that $c=1$.
Thus,  
\be
{U}_{\alpha\mu\nu}(q,r,p) = \underbrace{{\cal U}_{\alpha\mu\nu}(q,r,p)}_{{\cal O}(q^{0})}\,\, + \,\, 
\underbrace{{\cal U^{\,\prime}_{\alpha\mu\nu}}(q,r,p)}_{{\cal O}(q) }
\label{UUp}
\ee
Then, setting
\be
{\cal R}_{\alpha\mu\nu}(q,r,p) = {R}_{\alpha\mu\nu}(q,r,p) + {\cal U^{\,\prime}_{\alpha\mu\nu}}(q,r,p)
\label{RRp}
\ee
we arrive at the final separation 
\be
\label{Valtsep}
V_{\alpha\mu\nu}(q,r,p) = \underbrace{{\cal U}_{\alpha\mu\nu}(q,r,p)}_{m^2(q^2)} \,\, + \,\, 
\underbrace{{\cal R}_{\alpha\mu\nu}(q,r,p)}_{J_m(q^2)}
\ee
where the curly brackets below each term indicate 
to which equation [\1eq{Jeq} or \1eq{meq}] each term will contribute.


\section{\label{mbsf} Massless bound-state formalism}

Whereas in the SDE approach outlined in the previous section 
one relies predominantly on the global properties of the vertices $V$, 
within the massless bound-state formalism one 
takes, instead, a closer look at the field-theoretic  
composition of these vertices, establishing 
fundamental relations between their internal ingredients and the gluon mass.  
This becomes possible thanks to the 
key observation that, since  
the fully dressed vertices appearing in the diagrams of Fig.~\ref{QQ-SDE}  
are themselves governed by their own SDEs, 
the appearance of such massless poles  
must be associated with very concrete modifications in the 
various structures composing them. 

\begin{figure}[b!]
\center{\includegraphics[scale=0.6]{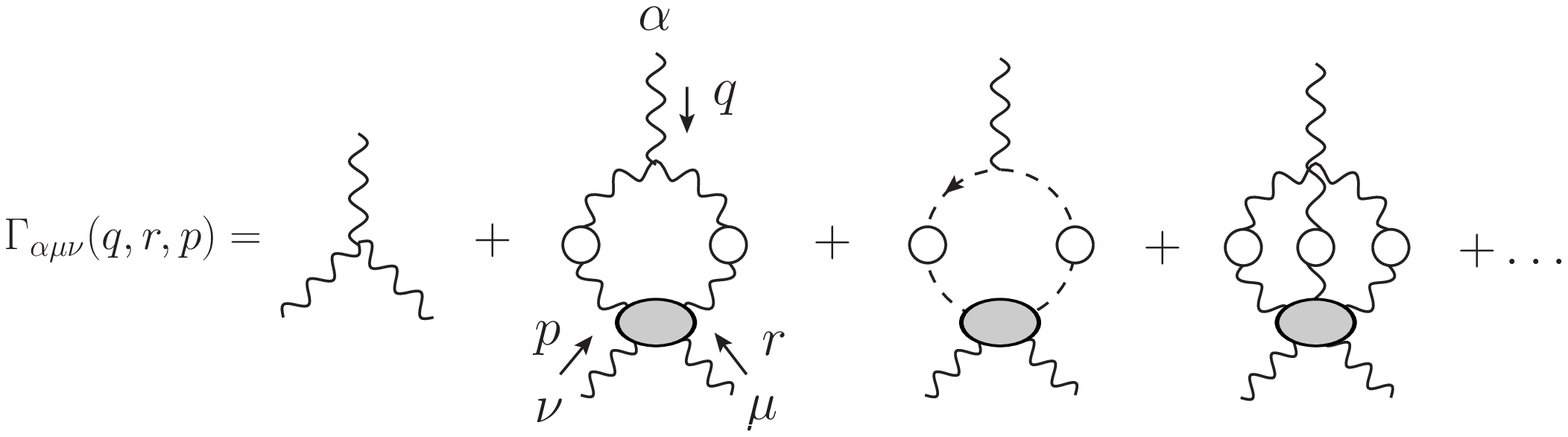}}
\caption{The SDE for the $Q^3$ vertex $\Gamma_{\alpha\mu\nu}(q,r,p)$.
Gray blobs denote the conventional 
1PI (with respect to vertical cuts) multiparticle 
kernels.}
\label{vertexSDE}
\end{figure}

Let us begin by recalling that, 
in general, when setting up the usual SDE for any fully-dressed 
vertex contained in \1eq{QQSDE}, a particular field (leg) 
is singled out, and is connected to the various multiparticle 
kernels through all elementary vertices of the theory involving 
this field (leg). The remaining legs enter into the various 
diagrams through the aforementioned multiparticle kernels, or, 
in terms of the standard skeleton expansions, through fully-dressed 
vertices (instead of tree-level ones). For example, in the case 
of the $Q^3$ three-gluon vertex $\Gamma_{\alpha\mu\nu}(q,r,p)$ 
we have (with a certain hindsight) identified the special leg to be the one 
entering into graph ($a_1$)  of Fig.~\ref{QQ-SDE} from the right, carrying momentum $q$; 
the corresponding vertex SDE is shown in Fig.~\ref{vertexSDE}. 

\begin{figure}[t!]
\center{\includegraphics[scale=0.5]{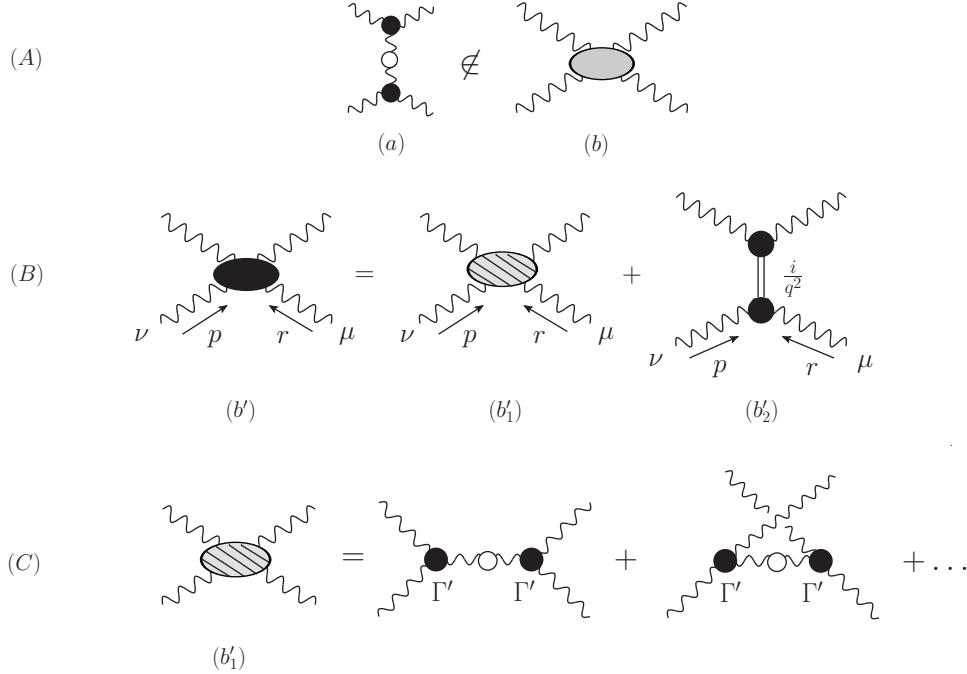}}
\caption{$(A)$ A diagram not included into the standard kernel. 
$(B)$ The kernel with the Schwinger mechanism turned on: in addition to the ``regular part'' $(b'_1)$ 
(gray striated), the massless excitation in the $q$-channel $(b'_2)$
is added. $(C)$ The part $(b'_1)$ is obtained from the 
original (gray) kernel $(b)$
by inserting massive gluon propagators into its diagrams, and carrying out the substitution 
\1eq{modifiedvertex} in the fully-dressed vertices of the skeleton expansion.}
\label{Modifiedkernel}
\end{figure}

Now, when the Schwinger mechanism is turned off, the various multiparticle 
kernels appearing in the SDE for the $Q^3$ vertex have a complicated 
skeleton expansion, but their common characteristic is that they are 
{\it one-particle irreducible} with respect to cuts in the direction of the 
momentum $q$. Thus, for example, diagram $(a)$ of Fig.~\ref{Modifiedkernel} 
is explicitly excluded from the (gray) four-gluon kernel $(b)$, and the same 
is true for all other kernels.

When the Schwinger mechanism is turned on, 
the structure of the kernels 
is modified by the presence of the composite massless excitations, 
described by a propagator of the type $i/q^2$. 
For example, as shown in Fig.~\ref{Modifiedkernel}, the gray four-gluon kernel 
is converted into a black kernel, diagram $(b')$, which is the sum of two parts:
\n{i} the term $(b'_1)$, which corresponds to a kernel (gray striated) that is 
``regular'' with respect to the $q$-channel, and 
\n{ii} the term $(b'_2)$, which describes the exchange of the composite 
massless excitation between two gluons in the $q$-channel.

\begin{figure}[t!]
\center{\includegraphics[scale=0.6]{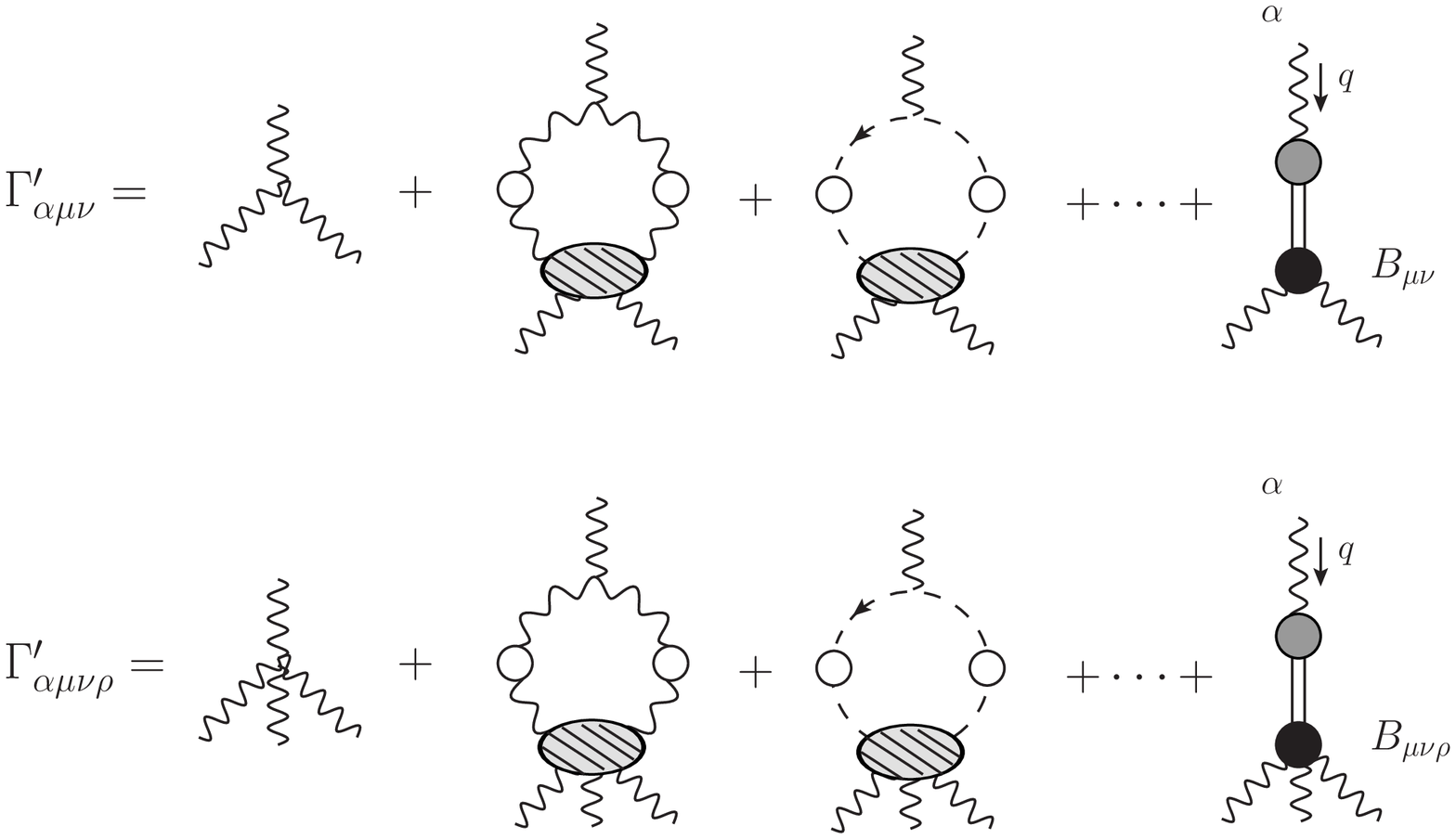}}
\caption{The SDEs for the full three and four gluon vertices in
 the presence of their pole parts. Note that $(i)$ the new SDE kernels 
 are modified with respect to those appearing in Fig.~\ref{vertexSDE}; $(ii)$ the last term in 
 these SDEs corresponds 
 to the ${\cal U}$ part of the pole vertices.}
\label{skeletonexpansions}
\end{figure}

Thus, when the  replacements of \1eq{massive} and \1eq{modifiedvertex} 
are carried out,  
the SDE for the different  interaction vertices in the presence  
of their pole parts  will  be given  by expansions  such  as 
those  shown  in Fig.~\ref{skeletonexpansions}.   

\begin{figure}[t!]
\center{\includegraphics[scale=0.6]{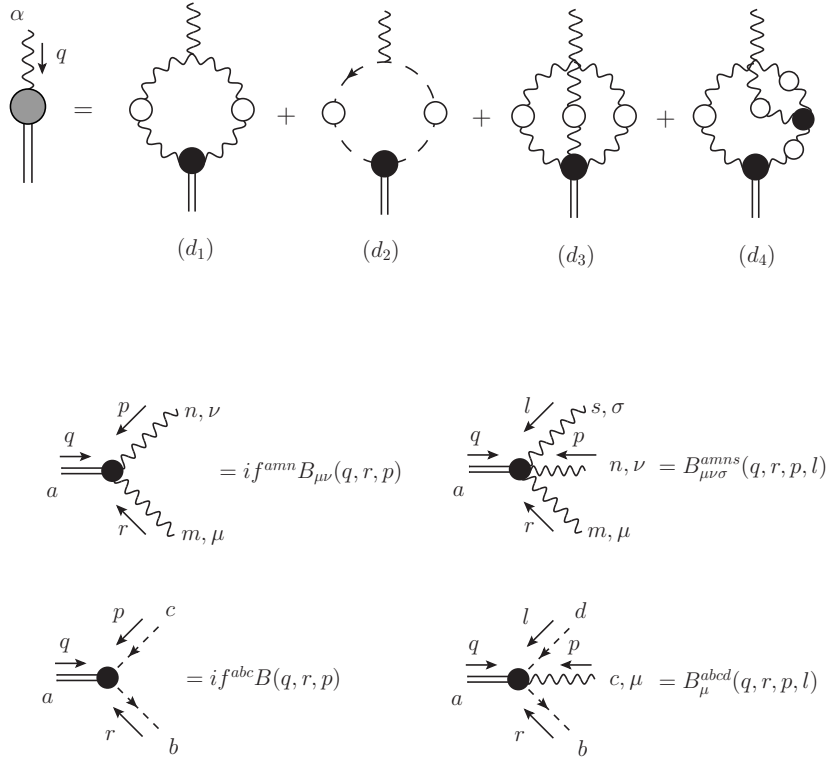}}
\caption{Diagrammatic representation of the transition amplitude $I_\alpha(q)$, 
and the Feynman rules for the different effective vertices $B$.}
\label{transitionandB}
\end{figure}

These modifications in the composition of the kernels give rise precisely to the 
vertices $V$ mentioned earlier. 
A closer look at the structure of the terms 
comprising the last term in Fig.~\ref{skeletonexpansions} reveals that the Lorentz index $\alpha$ (of the leg carrying the 
momentum $q$) is saturated precisely by the momentum $q$. Similarly, Bose symmetry 
forces the same behavior on the other two channels,
so that, in the end, we obtain a totally longitudinal structure, \ie  \1eq{totlon}. 

At this point it is advantageous to make the nonperturbative pole manifest, 
and cast the last term of the three-gluon vertex in
the form of Fig.~\ref{skeletonexpansions}, by setting
\begin{equation}\label{Ugenericthree}
{\cal U}_{\alpha\mu\nu}(q,r,p) = I_\alpha(q)\frac{i}{q^2}B_{\mu\nu}(q,r,p)\,;
\end{equation}
$I_\alpha(q)$ denotes the transition amplitude that mixes  a quantum gluon
with the massless excitation, $i/q^2$ corresponds to the propagator of
the massless excitation, and $B$  is an effective vertex describing the
interaction  between   the  massless  excitation   and  gluons  and/or
ghosts. In the standard language used in bound-state physics, 
$B$  represents the ``bound-state wave function'' (or ``Bethe-Salpeter wave function'').  
Clearly, due to Lorentz invariance,
\begin{equation}\label{Quantumtransition}
I_\alpha(q) = q_\alpha I(q^2),
\end{equation}
and the scalar cofactor, to be referred as the ``transition function'', 
is simply given by
\begin{equation}\label{scalarcofactor}
I(q^2) = \frac{q^\alpha}{q^2}I_\alpha(q).
\end{equation}
Furthermore, notice that the transition
amplitude $I_\alpha(q)$ is universal,  in the sense that it constitutes a common ingredient of 
all $V$ vertices, namely 
\begin{equation}\label{Ugeneric}
{\cal U}_{\alpha{\lbrace\dots\rbrace}}(q,\dots)=I_\alpha(q)\frac{i}{q^2} B_{{\lbrace\dots\rbrace}}(q,\dots);
\end{equation}
thus, the difference between the various ${\cal U}_{\alpha{\lbrace\dots\rbrace}}$ 
vertices is solely encoded into 
the structure of the wave functions $B_{{\lbrace\dots\rbrace}}$. 
The diagrammatic representation of the  transition amplitude 
as well as the different effective vertices $B$ are shown in Fig.~\ref{transitionandB}.
It is important to recognize that the ${\cal U}_{\alpha{\lbrace\dots\rbrace}}$ 
depend {\it quadratically} on the $B_{{\lbrace\dots\rbrace}}$, 
since $I_\alpha(q)$ depends linearly on them.

\begin{figure}[t!]
\center{\includegraphics[scale=0.6]{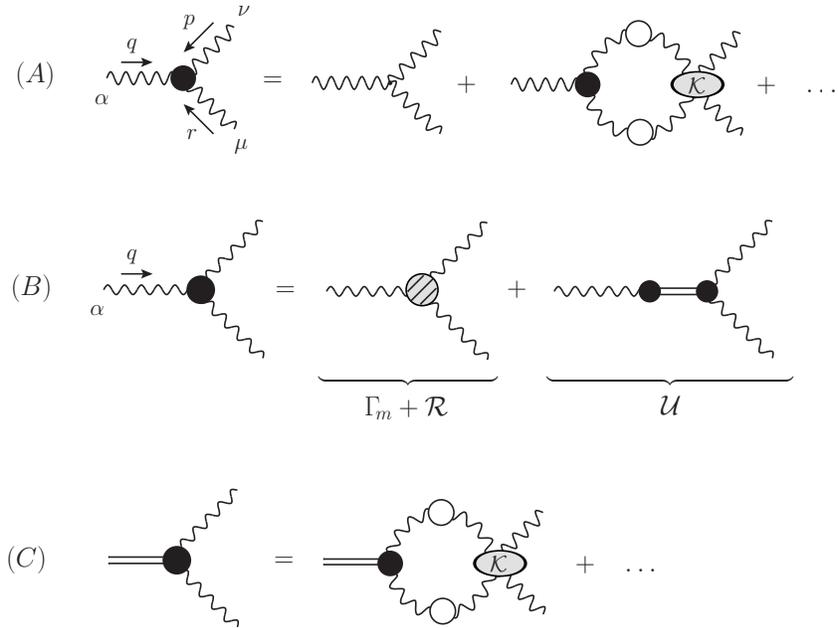}}
\caption{Basic steps in the derivation of the BSEs that govern the dynamics of the effective vertices $B$.}
\label{BSE}
\end{figure}

As has been explained in the literature, 
the dynamics of the ${B}$ functions may be determined, at least in principle,  
from a set of homogeneous (coupled and nonlinear) integral equations, 
known as BSEs. This particular set of equations 
must admit nontrivial solutions, 
which, when properly adapted to the kinematic details of the problem at hand, 
will furnish the momentum dependence of the wave functions ${B}$. 

The way to obtain this set of equations is by first rearranging the 
vertex SDE in such a way as to replace the bare vertex appearing in the 
original expansion ({\it e.g.},  Fig.~\ref{vertexSDE})
by a fully dressed one (Fig.~\ref{BSE}, line A).
At the same time, and as a consequence of  
this rearrangement, one 
must replace the standard SD kernel by the corresponding Bethe-Salpeter kernel
(denoted by ${\cal K}$ in Fig.~\ref{BSE}); 
the two kernels are diagrammatically different, and are formally related by 
a standard all-order formula~\cite{Maris:2003vk,LlewellynSmith:1969az,Bjorken:1979dk}.  
The next step is to separate the full vertex into the ``regular'' part, 
namely the part that behaves as a regular function in the limit  $q\to 0$, and the 
pole part, $1/q^2$, as shown in  Fig.~\ref{BSE}, line B. 
Note that the full regular part is the sum of $\Gamma\!_m$ and the term ${\cal R}$ coming from $V$, 
since neither of these two terms diverges in the aforementioned limit.
Then, this separation 
is carried out on both sides 
of the equation; since the vertex on the right-hand side (rhs) is now fully dressed, it too possesses a pole $1/q^2$.  
The final dynamical equation for ${B}$ emerges by equating 
the coefficients multiplying the pole term on both sides, as seen in Fig.~\ref{BSE}, line C. 
In the same way, the dynamical equation for the regular part is given by the remaining terms.

\section{\label{gmta} Relating the gluon mass with the transition amplitude}
The aim of this section is to derive the  fundamental formula that relates the 
effective gluon mass with the square of the transition amplitude.

To that end, let us go back to the SDE for the gluon propagator with the replacements given in 
Eqs.(\ref{modifiedvertex})-(\ref{massive}) already implemented, namely (Landau gauge)
\begin{equation}\label{massiveQQSDE}
[q^2J_m(q^2)-m^2(q^2)]P_{\mu\nu}(q) = q^2 P_{\mu\nu}(q)+i\sum_{i=1}^5(a'_i)_{\mu\nu},
\end{equation}

It turns out that 
the most expeditious way for deriving the gluon mass equation, and from it 
the desired relation between $m^2(q^2)$ and $I(q^2)$,  
is to identify, on both sides of \1eq{massiveQQSDE}, the 
cofactors of the tensorial structure $q_\mu q_\nu/q^2$ that survive the 
limit $q^2\rightarrow 0$, and then set them equal to each other. In doing so,  
it is clear that 
the left-hand side (lhs) of \1eq{massiveQQSDE} furnishes simply
\begin{equation}\label{masslhs}
[{\rm lhs}]_{\mu\nu} = \frac{q_\mu q_\nu}{q^2} m^2 (q^2).
\end{equation}
On the other hand, the corresponding contribution from the rhs is directly related to the 
$\cal U$ part of the $V$ vertices (see discussion in subsection \ref{strV}), which, 
due to their very definition [see \1eq{Ugeneric}], are all proportional to $q_{\nu}$.

\begin{figure}[t!]
\center{\includegraphics[scale=0.6]{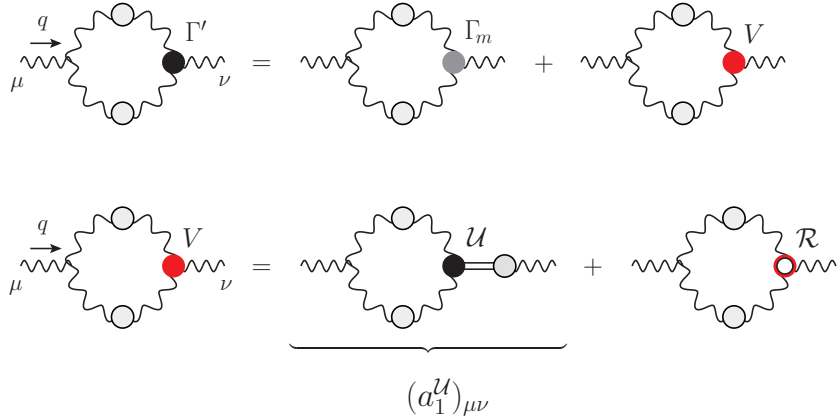}}
\caption{Procedure for isolating the contribution of diagram $(a'_1)$ to the gluon mass equation.}
\label{splitting}
\end{figure}

The procedure described above is exemplified in Fig.~\ref{splitting} for the particular 
case of the diagram $(a'_1)$.  
Specifically, in the first step one separates the regular part 
$\Gamma\!_m$ from the pole part $V$ of the full $Q^3$ vertex $\Gamma'$. 
In the second step the pole part $V$ is written as the sum of 
the ${\cal U}$ part (containing the explicit $q$-channel massless 
excitation) and the ${\cal R}$ part. 
Finally, due to the special structure of the ${\cal U}$ part,  
this contribution is proportional to $q_\mu q_\nu/q^2$, and contributes to the rhs of the mass equation.

\begin{figure}[t!]
\center{\includegraphics[scale=0.7]{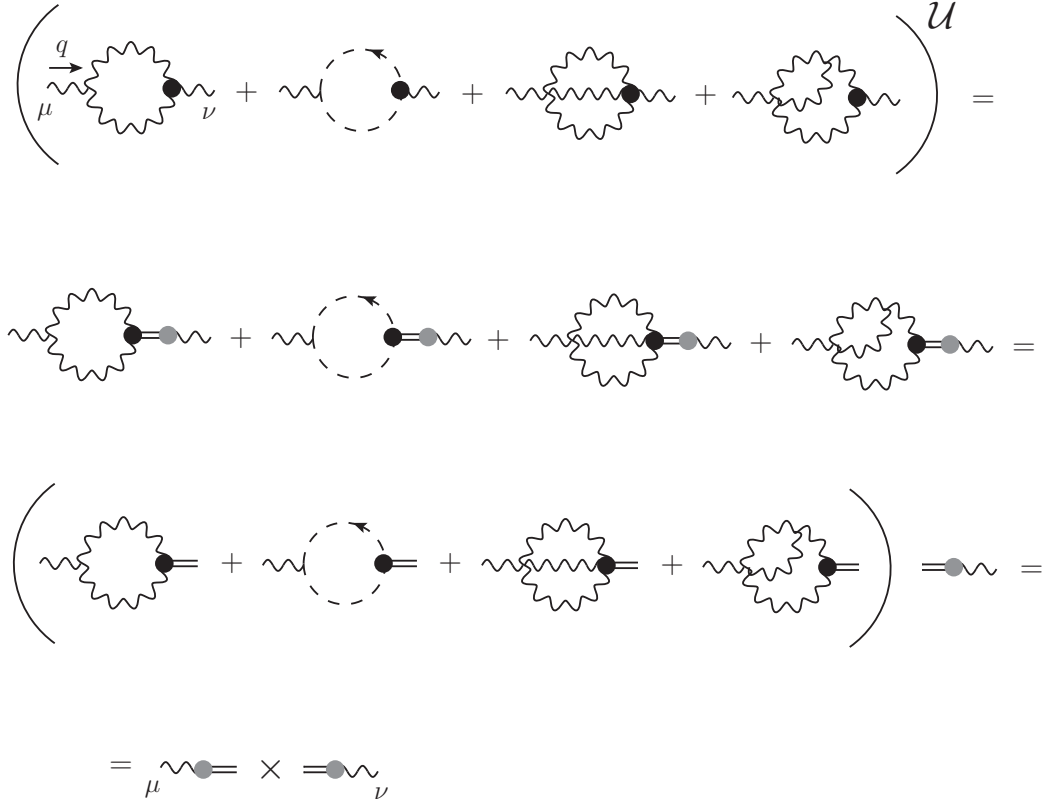}}
\caption{General procedure for determining the contributions of the rhs 
of \1eq{massiveQQSDE} to the effective gluon mass. Note 
that the vertices appearing in this figure correspond to the ${\cal U}$ 
parts of the full vertices $\Gamma'$.}
\label{QQmass}
\end{figure}

The next step is to carry out this procedure to the ${\cal U}$ parts of all diagrams, 
and determine the complete contribution to the rhs of the mass equation, 
as shown pictorially in Fig.~\ref{QQmass}.
It is clear that from all diagrams containing the ${\cal U}$ parts (first line of Fig.~\ref{QQmass}) one 
may factor out the common quantity $I(q^2)$, since, 
as we have emphasized in the previous section, $I(q^2)$ is universal  (second line of Fig.~\ref{QQmass}). 
Then, quite interestingly, the 
sum of the terms in the parenthesis is nothing else than the diagrammatic representation of  
$I(q^2)$, given in Fig.~\ref{transitionandB}
(note that all combinatorial factors work out exactly).

So, applying this procedure for each one of 
the fully-dressed vertices appearing in the SDE \1eq{massiveQQSDE}, one 
can put together the contributions of the several ${\cal U}$ parts 
to the gluon self-energy, as shown in Fig.~\ref{QQmass},
\begin{equation}
\sum_{i=1}^5(a^{{\cal U}}_i)_{\mu\nu} = g^2 I_\mu(q)\bigg(\frac{i}{q^2}\bigg) I_\nu(-q).
\label{PiU}
\end{equation}
Then, using \1eq{Quantumtransition}, together with the fact that $I_\nu(-q) = -I_\nu(q)$, 
one obtains the following result
\begin{equation}\label{massrhs}
[{\rm rhs}]_{\mu\nu} = \frac{q_\mu q_\nu}{q^2} g^2 I^2(q^2). 
\end{equation}
Therefore, equating \1eq{masslhs} with \1eq{massrhs} we find 
that the effective gluon mass is related to the transition 
amplitude through the simple formula (Minkowski space)
\begin{equation}\label{QQmassformula}
m^2(q^2) = g^2 I^2(q^2).
\end{equation}

This last formula may be passed to Euclidean space, using the standard 
conversion rules, together with the expression for $I(q^2)$ given in 
\1eq{transitionB1general}. In particular, the transition to the Euclidean 
space proceeds by using the standard formulas that allow the conversion of 
the various Green's functions from the Minkowski momentum $q^2$ to the Euclidean 
$q^2_{\chic E}=-q^2>0$; specifically
\begin{equation}\label{Euclideanrules}
\Delta_{\chic E}(q^2_{\chic E})=-\Delta(-q^2_{\chic E});\quad m_{\chic E}^2(q_{\chic E}^2)=m^2(-q_{\chic E}^2);\quad G_{\chic E}(q^2_{\chic E})=G(-q^2_{\chic E});\quad \int_k = i\int_{k_{\chic E}}.
\end{equation}
As a consequence, we have $I_{\chic E}(q^2_{\chic E})=-I(-q^2_{\chic E})$, 
so that 
\begin{equation}\label{QQmassformulaEuc}
m^2_{\chic E}(q^2_{\chic E}) = g^2 I_{\chic E}^2(q^2_{\chic E}).
\end{equation}

An immediate important implication of this last relation is that  
the gluon mass obtained is a positive-definite function for all values of the Euclidean momenta,
as expected on physical grounds.

Let us finally mention that the transversality of the full gluon self-energy [see \1eq{selfenmastran}]
guarantees that if one were to consider the part of the mass equation proportional to 
$g_{\mu\nu}$, one would eventually obtain exactly the same relation given in \1eq{QQmassformula}.
Note, however, that the corresponding derivation is far more subtle 
and laborious, and hinges crucially on the judicious use of a special integral identity 
(for a detailed treatment of this issue, see~\cite{Binosi:2012sj}).

\section{\label{bqi} The BQI of the transition amplitudes: SDE derivation}

We will now repeat the construction of the last section using the SDEs of the 
PT-BFM formalism. Specifically, 
we will derive the relations analogous to \1eq{QQmassformula}, which, in conjunction with the 
BQIs connecting the conventional and PT-BFM gluon propagators, will furnish a nontrivial relation between the 
corresponding transition amplitudes.

\subsection{General considerations}

In the PT-BFM formalism the natural separation of the gluonic field
into a ``quantum'' ($Q$) and a ``background'' ($B$) gives rise to an extended set of
Feynman rules, and leads to an increase in the type of possible Green's functions that 
one may consider. In the case of the gluonic two-point function,  
in addition to the conventional $QQ$ 
gluon propagator, $\Delta$, two additional quantities appear: 
the $QB$ propagator, $\widetilde{\Delta}$, 
mixing one quantum gluon with one background gluon, and the $BB$ propagator, 
$\widehat{\Delta}$, with two background gluon legs. It turns out that 
these three propagators are related by the all-order identities (referred to as BQIs)  
\begin{equation}
\Delta(q^2) = [1+G(q^2)]\widetilde{\Delta}(q^2) = [1+G(q^2)]^2\widehat{\Delta}(q^2).
\label{BQIpropagatorsa}
\end{equation}

\begin{figure}[!t]
\includegraphics[scale=.6]{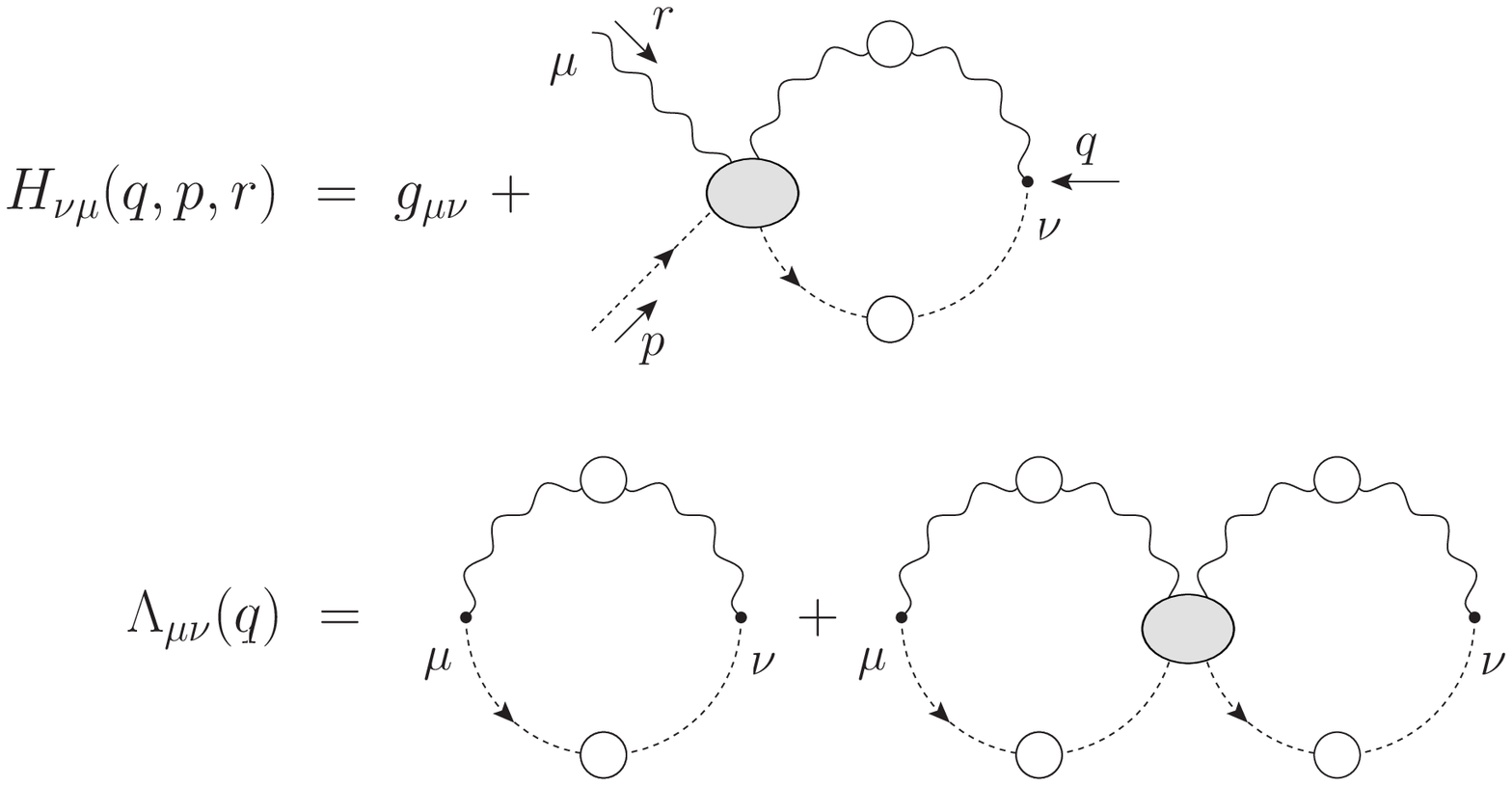}
\caption{\label{H-Lambda-new}Definitions and conventions of the auxiliary functions $\Lambda$ and $H$.}
\end{figure}

The function $G(q^2)$, whose role in enforcing these crucial relations is instrumental, 
is defined as the $g_{\mu\nu}$ form factor of a special two-point function, given by (see Fig.~\ref{H-Lambda-new})
\bea
\Lambda_{\mu\nu}(q)&=&-i\gA\int_k\!\Delta_\mu^\sigma(k)D(q-k)H_{\nu\sigma}(-q,q-k,k)\nonumber\\
&=&g_{\mu\nu}G(q^2)+\frac{q_\mu q_\nu}{q^2}L(q^2),
\label{Lambda}
\eea
where $C_A$ denotes the Casimir eigenvalue of the adjoint representation ($N$ for $SU(N)$), 
$d=4-\epsilon$ is the space-time dimension, and we have introduced the integral measure
\be
\int_{k}\equiv\frac{\mu^{\epsilon}}{(2\pi)^{d}}\!\int\!\mathrm{d}^d k,
\label{dqd}
\ee
with $\mu$ the 't Hooft mass. In addition, 
$D^{ab}(q^2)=\delta^{ab}D(q^2)$ is the ghost propagator, and $H_{\nu\sigma}$ is 
the gluon-ghost kernel~\cite{Pascual:1984zb,Davydychev:1996pb}.
The dressed loop expansion of $\Lambda$ and $H$ is shown in~\fig{H-Lambda-new}. 
Notice that the standard ghost-gluon vertex $\Gamma_{\mu}$ is obtained from  $H_{\nu\mu}$ simply through the contraction 
\be
q^{\nu} H_{\nu\mu}(q,p,r) = - \Gamma_{\mu}(r,q,p).
\label{ghH}
\ee
Finally, in the Landau gauge only, the form factors $G(q^2)$ and  $L(q^2)$ are linked to the 
ghost dressing function 
\be
F(q^2)=q^2D(q^2)
\ee
by means of the all-order relation~\cite{Grassi:2004yq,Aguilar:2009nf,Aguilar:2009pp,Aguilar:2010gm}  
\be
F^{-1}(q^2) = 1 + G(q^2) + L(q^2) \,,
\label{funrel}
\ee
which will be extensively used in the ensuing analysis. 

Returning to \1eq{BQIpropagatorsa}, it is important to recognize that the two 
basic functions constituting the gluon propagators, namely 
$\Jm(q^2)$and $m^2(q^2)$, satisfy \1eq{BQIpropagatorsa} individually~\cite{Aguilar:2011ux}. In particular, the 
corresponding masses are related by 
\begin{equation}\label{BQImasses}
\widehat{m}^2(q^2) = [1+G(q^2)]\widetilde{m}^2(q^2) = [1+G(q^2)]^2 m^2(q^2).
\end{equation}

\begin{figure}[t]
\center{\includegraphics[scale=0.6]{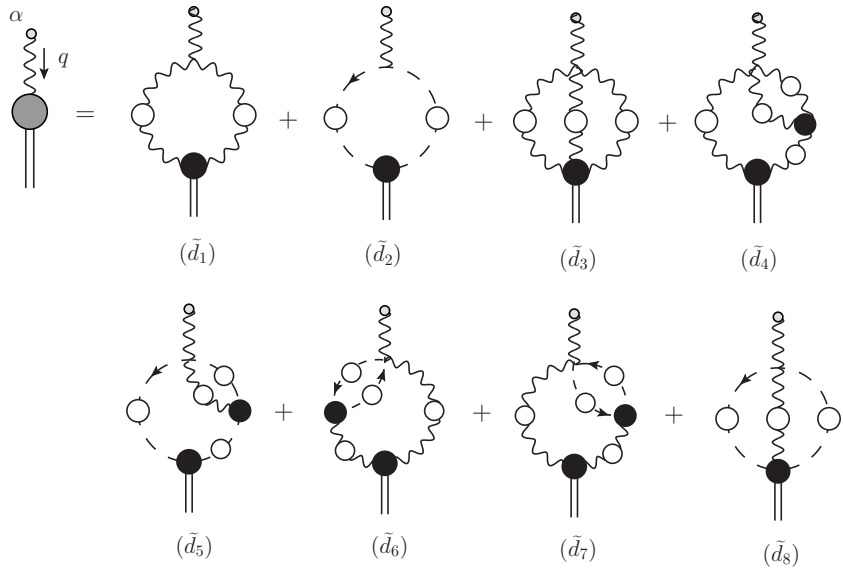}}
\caption{Diagrammatic representation of the background transition amplitude ${\widetilde I}_\alpha(q)$.}
\label{BFMtransition}
\end{figure}

Finally, in order to obtain from the SDEs of the PT-BFM propagators
a mass formula analogous to that of \1eq{QQmassformula}, 
one needs to introduce the appropriate transition amplitude  
connecting the background gluon with the massless excitation. This new transition 
amplitude, whose diagrammatic representation is shown in Fig.~\ref{BFMtransition},
allows us to write the $\widetilde{\cal U}$ part of the corresponding pole 
vertices in the BFM as
\begin{equation}\label{backgroundUgeneric}
\widetilde{\cal U}_{\alpha{\lbrace\dots\rbrace}}(q,\dots)=\widetilde{I}_\alpha(q)\frac{i}{q^2}B_{{\lbrace\dots\rbrace}}(q,\dots).
\end{equation}
Observe that 
only the transition amplitude is modified in this expression 
with respect to \1eq{Ugeneric} when we go to the BFM. 
This is so because the only background field is the one 
carrying the momentum $q$, while all other fields 
are quantum (\ie they are common to the PT-BFM and conventional descriptions).

\subsection{Relating the transition amplitudes through the SDEs of the PT-BFM}

\begin{figure}[t]
\center{\includegraphics[scale=0.45]{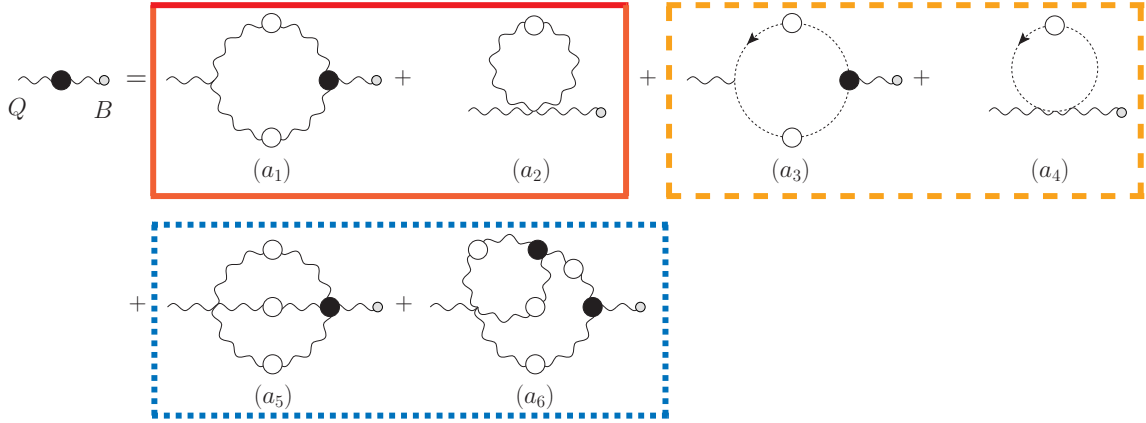}}
\caption{SDE obeyed by the gluon self-energy with one 
quantum ($Q$) and one background gluon ($B$) entering. Each box encloses a set of diagrams 
forming a transverse subgroup. The small gray circles appearing on the external 
legs (entering from the right, only!) are used to indicate background gluons.}
\label{QB-SDE}
\end{figure}

Consider the SDE for the $QB$ propagator, connecting 
a quantum and a background field, given by the expression
\begin{equation}\label{QBSDE}
\widetilde\Delta^{-1}(q^2) P_{\mu\nu}(q) = q^2P_{\mu\nu}(q) + i\sum_{i=1}^6 (a_i)_{\mu\nu},
\end{equation}
with the diagrams $(a_i)$ shown in Fig.~\ref{QB-SDE}. 
Note that the diagrams are separated into three blocks, 
each of which is individually transverse; this special 
property of ``block-wise'' transversality is particular to the PT-BFM scheme.

Let us now apply the diagrammatic procedure shown in Fig.~\ref{QQmass} to the new 
set of diagrams appearing in this SDE. Evidently, the quantity to be factored out from all diagrams 
comprising the $\widetilde{\cal U}$-related part of the SDE is  $\widetilde{I}(q^2)$. 
Then, it is easy to 
recognize that the 
(sub) diagrams composing the other factor coincide precisely with those shown 
in the first line of Fig.~\ref{transitionandB}, 
thus giving rise to $I(q^2)$ again. 
Thus we arrive at the following expression for the mass of the $QB$ propagator,
\begin{equation}\label{QBmassformula}
\widetilde{m}^2(q^2) = g^2 I(q^2)\widetilde{I}(q^2).
\end{equation} 

At this point one may use the BQI of \1eq{BQImasses} 
to replace $\widetilde{m}^2(q^2)$ in favor of $m^2(q^2)$
on the lhs of \1eq{QBmassformula}, namely
\begin{equation}\label{subs}
m^2(q^2) = \frac{g^2 I(q^2) \widetilde{I}(q^2)}{1+G(q^2)}.
\end{equation}
Note that this last substitution is legitimate, because the corresponding 
SDEs have been considered in their full, untruncated version (all diagrams kept); therefore 
the masses obtained from them are the same exact quantities that satisfy the BQI. 

Then, direct comparison with \1eq{QQmassformula} furnishes the central relation 
\begin{equation}\label{BQItransition}
\widetilde{I}(q^2) = [1 + G(q^2)] I(q^2).
\end{equation}
Interestingly enough, this relation emerges as a self-consistency requirement between the 
results obtained from two formally different, but physically equivalent,
versions of the gluon propagator SDE.  

Furthermore, due to the special 
structure of \1eq{backgroundUgeneric}, where only the transition amplitude knows if the $q$-leg 
is quantum or background, the result \1eq{BQItransition} implies that the part $\widetilde{\cal U}$ 
of the pole vertices must also satisfy the same BQI, namely (suppressing indices)
\begin{equation}\label{BQIU}
\widetilde{\cal U}=[1+G(q^2)]{\cal U}.
\end{equation}

\begin{figure}[t]
\center{\includegraphics[scale=0.45]{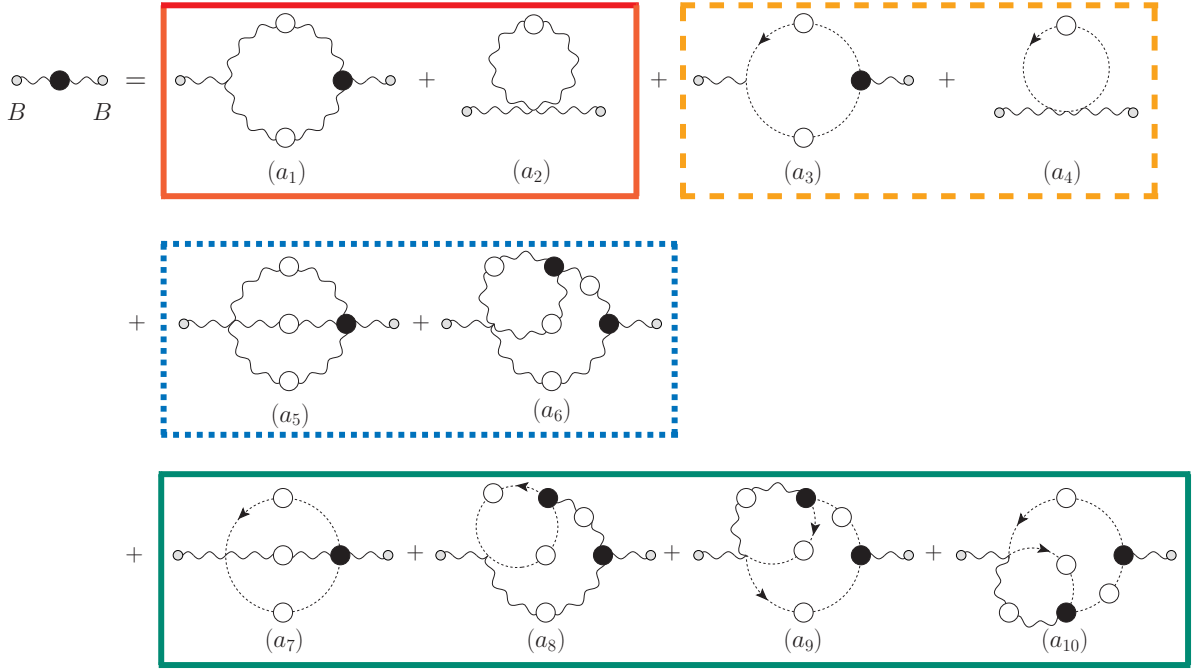}}
\caption{SDE obeyed by the gluon self-energy with two background gluons ($B$) 
entering. The graphs inside each box form a transverse subgroup.
External background legs are indicated by the small gray circles.}
\label{BB-SDE}
\end{figure}

Finally, let us consider for completeness 
the SDE for the $BB$ propagator connecting two background 
gluons, given by the expression
\begin{equation}\label{BBSDE}
\widehat\Delta^{-1}(q^2) P_{\mu\nu}(q) = q^2P_{\mu\nu}(q) + i\sum_{i=1}^{10} (a_i)_{\mu\nu},
\end{equation}
where the diagrams $(a_i)$ appearing on the rhs of the 
equation are shown in Fig.~\ref{BB-SDE}, 
arranged again into individually transverse blocks.

Repeating the procedure of Fig.~\ref{QQmass} for this last SDE, we see that, as in the $QB$ case,   
the common quantity to be factored out from all diagrams 
is  again $\widetilde{I}(q^2)$. Then, the corresponding sum of (sub) diagrams 
coincides precisely with those of 
Fig.~\ref{BFMtransition}, thus giving rise to another $\widetilde{I}(q^2)$. As a result, one obtains 
\begin{equation}\label{BBmassformula}
\widehat{m}^2(q^2) = g^2\widetilde{I}^2(q^2).
\end{equation}
It should be easy to verify at this point that the direct use of \1eq{BQImasses}  
and subsequent comparison with \1eq{QQmassformula} [or with \1eq{QBmassformula}], 
furnishes again the result of \1eq{BQItransition}.

\section{\label{excon} Three-gluon pole vertex}
In the previous section the basic relations \1eq{BQItransition} and
\1eq{BQIU} have been derived at the level of the SDEs, by appealing to
the general properties  of the massless bound state  formalism and the
BQIs relating the gluon propagators in the PT-BFM formalism.  
In  the next  two subsections  we will  derive  the same
relations from  the pole parts  of the $Q^3$ vertex
and the  $BQ^2$ vertex.  Specifically, in  the first subsection we  derive in detail
the closed form  of these vertices, 
while in the  next we identify 
their ${\cal U}$ parts, and carry out a direct comparison.

\subsection{\label{expcon}Explicit construction}

As has been explained in detail in the early~\cite{Jackiw:1973tr,Jackiw:1973ha,Cornwall:1973ts,Eichten:1974et,Poggio:1974qs}
and recent literature~\cite{Binosi:2012sj,Ibanez:2011hc}, 
in order to preserve the gauge invariance of the theory in the presence of 
gauge boson masses, the usual vertices must be supplemented by pole parts.
In particular, the pole part of the $BQ^2$ vertex must satisfy the following WI and STIs~\cite{Binosi:2012sj,Ibanez:2011hc},
\begin{eqnarray}\label{poleBQQSTI}
&& q^\alpha\widetilde{V}_{\alpha\mu\nu}(q,r,p)= m^2(r^2)P_{\mu\nu}(r)-m^2(p^2)P_{\mu\nu}(p), \nonumber\\
&& r^\mu\widetilde{V}_{\alpha\mu\nu}(q,r,p)= F(r^2)[m^2(p^2)P_\nu^\mu(p)\widetilde{H}_{\mu\alpha}(p,r,q)-\widetilde{m}^2(q^2)P_\alpha^\mu(q)H_{\mu\nu}(q,r,p)], \nonumber\\
&& p^\nu\widetilde{V}_{\alpha\mu\nu}(q,r,p)= F(p^2)[\widetilde{m}^2(q^2)P_\alpha^\nu(q)H_{\nu\mu}(q,p,r)-m^2(r^2)P_\mu^\nu(r)\widetilde{H}_{\nu\alpha}(r,p,q)].
\end{eqnarray}
The quantity $\widetilde{H}$ is given by the same diagrammatic representation 
as that of $H$, shown in \fig{H-Lambda-new},  
but with the incoming gluon field replaced by a background one.

In the pole part of the $Q^3$ vertex, the leg associated with $q_\alpha$ is quantum instead of background, 
and, as a consequence, the Abelian-like WI [first in \1eq{poleBQQSTI}] is replaced by an 
STI, namely
\begin{equation}\label{poleQQQSTI}
q^\alpha V_{\alpha\mu\nu}(q,r,p)=F(q^2)[m^2(r^2)P_\mu^\alpha(r)H_{\alpha\nu}(r,q,p)-m^2(p^2)P_\nu^\alpha(p)H_{\alpha\mu}(p,q,r)].
\end{equation}
The STIs with respect to the other two legs are those of \1eq{poleBQQSTI}, 
but with the ``tilded'' quantities replaced by conventional ones. 

It turns out that the explicit closed form of the two pole vertices in question, namely $\widetilde{V}$ 
and $V$, may be determined 
from the STIs they satisfy, and the requirement of complete longitudinality, 
\ie the condition \1eq{totlon}.
Specifically, opening up the transverse projectors in \1eq{totlon}, one can 
write the entire vertex in terms of its own divergences,
\begin{eqnarray}\label{termsV}
\widetilde{V}_{\alpha\mu\nu}(q,r,p) &=& \frac{q_\alpha}{q^2}q^\beta\widetilde{V}_{\beta\mu\nu} + \frac{r_\mu}{r^2}r^\rho\widetilde{V}_{\alpha\rho\nu} + \frac{p_\nu}{p^2}p^\sigma\widetilde{V}_{\alpha\mu\sigma} - \frac{q_\alpha r_\mu}{q^2r^2}q^\beta r^\rho \widetilde{V}_{\beta\rho\nu} - \frac{q_\alpha p_\nu}{q^2p^2}q^\beta p^\sigma \widetilde{V}_{\beta\mu\sigma} \nonumber\\
&-& \frac{r_\mu p_\nu}{r^2p^2}r^\rho p^\sigma \widetilde{V}_{\alpha\rho\sigma} + \frac{q_\alpha r_\mu p_\nu}{q^2r^2p^2}q^\beta r^\rho p^\sigma \widetilde{V}_{\beta\rho\sigma}.
\end{eqnarray}
Note that the last term will not contribute because if we apply the STI's,
\begin{equation}\label{zerotermV}
q^\beta r^\rho p^\sigma \widetilde{V}_{\beta\rho\sigma}(q,r,p)=0.
\end{equation}
So, using \1eq{poleBQQSTI} to evaluate the various terms, and after a 
straightforward rearrangement, 
we obtain the following expression for the pole part of the $BQ^2$ vertex,
\begin{eqnarray}\label{nonBoseBQQV}
\widetilde{V}_{\alpha\mu\nu}(q,r,p) &=& \frac{q_\alpha}{q^2}[m^2(r^2)-m^2(p^2)]P_\mu^\rho(r)P_{\rho\nu}(p) \nonumber\\
&+& D(r^2)[m^2(p^2)P_\nu^\rho(p)\widetilde{H}_{\rho\alpha}(p,r,q)-\widetilde{m}^2(q^2)P_\alpha^\rho(q)P_\nu^\sigma(p)H_{\rho\sigma}(q,r,p)]r_\mu \nonumber\\
&+& D(p^2)[\widetilde{m}^2(q^2)P_\alpha^\rho(q)H_{\rho\mu}(q,p,r)-m^2(r^2)P_\mu^\rho(r)\widetilde{H}_{\rho\alpha}(r,p,q)]p_\nu.
\end{eqnarray}

Applying the same procedure, but using now the STIs of \1eq{poleQQQSTI}, 
together with the condition of \1eq{totlon}, 
we derive the closed expression for the pole part of the $Q^3$ vertex, 
\begin{eqnarray}\label{nonBoseQQQV}
V_{\alpha\mu\nu}(q,r,p) &=& D(q^2)[m^2(r^2)H_{\rho\sigma}(r,q,p)-m^2(p^2)H_{\sigma\rho}(p,q,r)]P_\mu^\rho(r)P_\nu^\sigma(p)q_\alpha \nonumber\\
&+& D(r^2)[m^2(p^2)P_\nu^\rho(p)H_{\rho\alpha}(p,r,q)-m^2(q^2)P_\alpha^\rho(q)P_\nu^\sigma(p)H_{\rho\sigma}(q,r,p)]r_\mu \nonumber\\
&+& D(p^2)[m^2(q^2)P_\alpha^\rho(q)H_{\rho\mu}(q,p,r)-m^2(r^2)P_\mu^\rho(r)H_{\rho\alpha}(r,p,q)]p_\nu.
\end{eqnarray}

Let us discuss next certain issues related to the self-consistency of
the previous vertex  construction. Notice that, in order to obtain the expressions in 
\1eq{nonBoseBQQV} and \1eq{nonBoseQQQV}, one must apply sequentially the WI and the
STIs. In  doing so, the Bose  symmetry of both vertices  is no longer
explicit, and  the result obtained  is not manifestly  symmetric under
the  exchange of the quantum  gluon legs.  Furthermore, seemingly  different
expressions are obtained,  depending on which of the  two momenta acts
first  on $\widetilde{V}$ or $V$.  However,  if  one   imposes  the  simple
requirement of  algebraic commutativity between  the WI and  the STIs
satisfied  by  the  three-gluon  vertex,  the  Bose  symmetry  becomes
manifest.  For  example,  using   \1eq{nonBoseBQQV}  we  can  see  that  the
elementary requirement
\begin{equation}\label{commutative}
q^\alpha r^\mu \widetilde{V}_{\alpha\mu\nu}(q,r,p) = r^\mu q^\alpha \widetilde{V}_{\alpha\mu\nu}(q,r,p),
\end{equation}
gives raise to the following identity
\begin{equation}\label{strangeidentity1}
F(r^2)P_\nu^\mu(p)q^\alpha \widetilde{H}_{\mu\alpha}(p,r,q) = -r_\mu P_\nu^\mu(p).
\end{equation}
A similar identity is obtained by imposing the requirement of \1eq{commutative} 
at the level of $V$, namely
\begin{equation}\label{strangeidentity2}
F(r^2)P_\nu^\mu(p)q^\alpha H_{\mu\alpha}(p,r,q) = -F(q^2)P_\nu^\mu(p)r^\alpha H_{\mu\alpha}(p,q,r).
\end{equation}
Quite interestingly, the identities in \1eq{strangeidentity1} and \1eq{strangeidentity2} are a direct 
consequence of the WI and the STI that the kernels $H$ and $\widetilde{H}$ satisfy, 
when they are contracted with the momentum of the background or quantum gluon leg, namely~\cite{Binosi:2011wi}
\begin{eqnarray}\label{STIH}
&& q^\alpha\widetilde{H}_{\mu\alpha}(p,r,q) = -p_\mu F^{-1}(p^2)-r_\mu F^{-1}(r^2), \nonumber\\
&& q^\alpha H_{\mu\alpha}(p,r,q) = -F(q^2)[p_\mu F^{-1}(p^2)C(p,q,r)+r^\alpha F^{-1}(r^2) H_{\mu\alpha}(p,q,r)],
\end{eqnarray}
where $C(p,q,r)$ is an auxiliary ghost function.

Indeed, use of \1eq{STIH} into \1eq{strangeidentity1} and \1eq{strangeidentity2}, respectively, 
leads to a trivial identity. 
Conversely, one may actually derive \1eq{STIH} from \1eq{strangeidentity1} and \1eq{strangeidentity2}; 
for example, starting with \1eq{strangeidentity1}, and using also the identities~\cite{Binosi:2011wi} 
\begin{eqnarray}\label{STIHandGamma}
&& p^\mu \widetilde{H}_{\mu\alpha}(p,r,q) = r_\alpha F^{-1}(r^2) - \widetilde{\Gamma}_\alpha(q,p,r), \nonumber\\
&& q^\alpha \widetilde{\Gamma}_\alpha(q,p,r) = p^2 F^{-1}(p^2)-r^2 F^{-1}(r^2),
\end{eqnarray}
one can easily reproduce \1eq{STIH}.

Evidently, these constraints allow us to cast the pole part of the $BQ^2$ vertex into a manifestly Bose symmetric form with respect to the quantum legs,
\begin{equation}\label{BoseBQQV}
\widetilde{V}_{\alpha\mu\nu}(q,r,p) = \frac{q_\alpha}{q^2}[m^2(r^2)-m^2(p^2)]P_\mu^\rho(r)P_{\rho\nu}(p) + 
\widetilde{S}_{\alpha\mu\nu}(q,r,p) - \widetilde{S}_{\alpha\nu\mu}(q,p,r),
\end{equation}
with
\begin{eqnarray}\label{BosetermBQQV}
\widetilde{S}_{\alpha\mu\nu}(q,r,p) &=& D(r^2)m^2(p^2)P_\nu^\rho(p)\widetilde{H}_{\rho\alpha}(p,r,q)r_\mu \nonumber\\
&-& \frac{r_\mu}{2}D(r^2)\widetilde{m}^2(q^2)P_\alpha^\rho(q)[g_\nu^\sigma + P_\nu^\sigma(p)]H_{\rho\sigma}(q,r,p).
\end{eqnarray}
Finally, for the pole part of the $Q^3$ vertex, the Bose symmetric expression reads
\begin{equation}\label{BoseQQQV}
V_{\alpha\mu\nu}(q,r,p) = S_{\alpha\mu\nu}(q,r,p) - S_{\mu\alpha\nu}(r,q,p) - S_{\nu\mu\alpha}(p,r,q),
\end{equation}
with
\begin{eqnarray}\label{BosetermQQQV}
S_{\alpha\mu\nu}(q,r,p) &=& \frac{q_\alpha}{2}D(q^2)m^2(r^2)P_\mu^\rho(r)[g_\nu^\sigma + P_\nu^\sigma(p)]H_{\rho\sigma}(r,q,p) 
\nonumber\\
&-& \frac{q_\alpha}{2}D(q^2)m^2(p^2)P_\nu^\rho(p)[g_\mu^\sigma + P_\mu^\sigma(r)]H_{\rho\sigma}(p,q,r).
\end{eqnarray}
Note the fact that, as it should be, setting in Eqs.(\ref{BoseBQQV})-(\ref{BoseQQQV}) the tree-level values $F=1$ and $H_{\mu\nu}=g_{\mu\nu}$ for the ghost dressing function and the gluon-ghost kernel, respectively, one recovers the expression for the ``abelianized'' three-gluon vertex first presented in~\cite{Cornwall:1985bg}.

\subsection{\label{upart} Transition BQI derived from the pole vertices}

We next identify from the explicit expressions for the $BQ^2$ and $Q^3$ pole vertices 
the terms $\widetilde{{\cal U}}$ and ${\cal U}$. 
To that end, and in complete accordance with the discussion presented in subsection \ref{strV},  
we apply a kinematic and a dynamical criterion, namely 
\n{i} determine $U$ by collecting all  terms containing the tensorial structure 
$q_\alpha/q^2$, and \n{ii} extract ${\cal U}$ by discarding 
(and reassigning them to ${\cal U^{\,\prime}}$) 
all terms that, when inserted into the SDE of
the gluon  self-energy, do not survive the $q\rightarrow 0$  limit, \ie do not contribute to the gluon mass equation.
The final objective is to infer the validity of the BQI of \1eq{BQIU}, 
connecting $\widetilde{{\cal U}}$ and ${\cal U}$. 

Applying criterion \n{i}, and denoting by  ${\widetilde U}$ and $U$ the resulting expressions,
it is straightforward to obtain from \1eq{BoseBQQV}
\begin{eqnarray}\label{BQQUexpression}
\widetilde{U}_{\alpha\mu\nu}(q,r,p) &=& \frac{q_\alpha}{q^2}\bigg\lbrace m^2(r^2)P_\mu^\rho(r)P_{\rho\nu}(p) - \frac{r_\mu}{2}D(r^2)\widetilde{m}^2(q^2)[g_\nu^\rho + P_\nu^\rho(p)]\Gamma_\rho(p,q,r)\bigg\rbrace \nonumber\\
&-& \binom{r\leftrightarrow p}{\mu \leftrightarrow \nu},
\end{eqnarray}
where \1eq{ghH} has been applied in order to relate the contractions of the auxiliary ghost 
functions $H$ with the conventional gluon-ghost vertices $\Gamma_\rho$, and 
$\binom{r\leftrightarrow p}{\mu \leftrightarrow \nu}$ is obtained from the term that is shown explicitly after carrying out the 
exchanges indicated. In addition, for the $\widetilde{R}$ part of the vertex we find 
\begin{eqnarray}\label{BQQRexpression}
\widetilde{R}_{\alpha\mu\nu}(q,r,p) &=& \frac{r_\mu}{r^2}F(r^2)\bigg\lbrace m^2(p^2)P_\nu^\rho(p)\widetilde{H}_{\rho\alpha}(p,r,q) - \frac{1}{2}\widetilde{m}^2(q^2)[g_\nu^\rho + P_\nu^\rho(p)] H_{\alpha\rho}(q,r,p)\bigg\rbrace \nonumber \\
&-& \binom{r\leftrightarrow p}{\mu \leftrightarrow \nu}.
\end{eqnarray}
Similarly, from \1eq{BoseQQQV}, and using the special STI \1eq{STIH}, we obtain 
\begin{eqnarray}\label{stillnot}
U_{\alpha\mu\nu}(q,r,p) &=& \frac{q_\alpha}{q^2}\bigg\lbrace F(q^2) m^2(r^2)H_{\rho\sigma}(r,q,p)P_\mu^\rho(r)P_\nu^\sigma(p) \nonumber\\
&-& \frac{r_\mu}{2}D(r^2)m^2(q^2)[g_\nu^\rho + P_\nu^\rho(p)] \Gamma_\rho(p,q,r) \bigg\rbrace - \binom{r\leftrightarrow p}{\mu \leftrightarrow \nu},
\end{eqnarray}
and
\begin{eqnarray}\label{QQQRexpression}
R_{\alpha\mu\nu}(q,r,p) &=& \frac{r_\mu}{r^2}F(r^2)\bigg\lbrace m^2(p^2)P_\nu^\rho(p)H_{\rho\alpha}(p,r,q) - \frac{1}{2}m^2(q^2)[g_\nu^\rho + P_\nu^\rho(p)] H_{\alpha\rho}(q,r,p)\bigg\rbrace \nonumber \\
&-& \binom{r\leftrightarrow p}{\mu \leftrightarrow \nu}.
\end{eqnarray}

Let us now assume for a moment that 
$\widetilde{{U}} = \widetilde{{\cal U}}$ 
and $U= {\cal U}$, as well as $\widetilde{R} = \widetilde{{\cal R}}$ and $R = {\cal R}$. 
Let us further use \1eq{BQImasses}, and substitute  $m^2(q^2) = \widetilde{m}^2(q^2) [1+G(q^2)]^{-1}$ into 
\1eq{stillnot}. Then, employing \1eq{funrel} valid in the Landau gauge, dropping $L(q^2)$ 
since $L(0)=0$ [thus applying effectively criterion (ii)], it is relatively easy to  
establish that the (would-be) $\widetilde{{\cal U}}$ and ${\cal U}$ fail to satisfy  \1eq{BQIU}
due to the presence of a very particular term. Specifically, if  $H_{\rho\sigma} \to g_{\rho\sigma}$ in the first line 
of \1eq{stillnot}, then the BQI of \1eq{BQIU} would be satisfied.

\begin{figure}[ht]
\center{\includegraphics[scale=0.5]{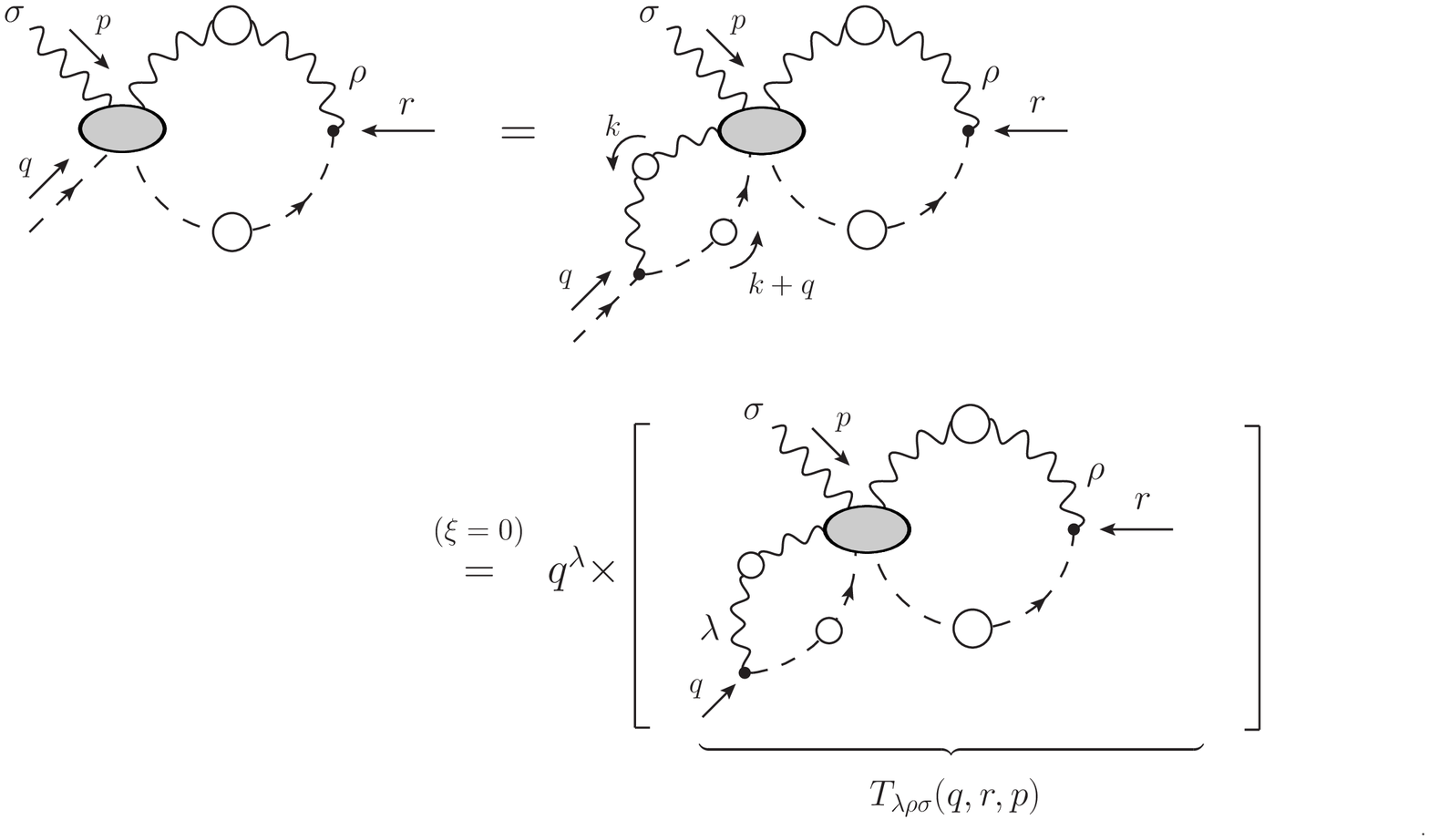}}
\caption{Diagrammatic representation of \1eq{quantumHT}.}
\label{Hquantum}
\end{figure}

Therefore, write the $H_{\rho\sigma}(r,q,p)$ in \1eq{stillnot} as 
\begin{equation}\label{splitH}
H_{\rho\sigma}(r,q,p) = g_{\rho\sigma} + H^Q_{\rho\sigma}(r,q,p),
\end{equation}
where the $g_{\rho\sigma}$ corresponds to the tree-level value of $H_{\rho\sigma}$, and $H^Q_{\rho\sigma}$ 
contains the all-order quantum corrections;
an exactly analogous expression holds for $H_{\sigma\rho}(p,q,r)$, with $p\leftrightarrow r$.
Note that in both cases the momentum $q$ is carried by the ghost leg, see Fig.~\ref{Hquantum}.
Then, if we define  the part ${\cal U}$ as [see \1eq{UUp}]
\begin{equation}\label{trueU}
{\cal U}_{\alpha\mu\nu}(q,r,p) = U_{\alpha\mu\nu}(q,r,p) - {\cal U^{\,\prime}_{\alpha\mu\nu}}(q,r,p), 
\end{equation}
with 
\begin{equation}\label{badterm}
{\cal U^{\,\prime}_{\alpha\mu\nu}}(q,r,p) = \frac{q_\alpha}{q^2}F(q^2)m^2(r^2)P_\mu^\rho(r)P_\nu^\sigma(p)H_{\rho\sigma}^Q(r,q,p) - \binom{r\leftrightarrow p}{\mu \leftrightarrow \nu}\,,
\end{equation}
it is clear that 
\begin{equation}\label{threegluonBQIU}
\widetilde{{\cal U}}_{\alpha\mu\nu}(q,r,p) = [1+G(q^2)] {\cal U}_{\alpha\mu\nu}(q,r,p).
\end{equation}
 
The real justification for 
discarding ${\cal U^{\,\prime}_{\alpha\mu\nu}}$ from $U_{\alpha\mu\nu}$ is 
provided precisely by invoking criterion \n{ii}: 
the discarded term does not survive the limit $q\rightarrow 0$ when inserted into the 
corresponding gluon SDE.

To demonstrate that this is indeed so, 
note that in the Landau gauge 
\mbox{$(k+q)^\lambda\Delta_{\lambda\sigma}(k) = q^\lambda\Delta_{\lambda\sigma}(k)$}, 
and the momentum of the incoming ghost leg factorizes out of the loop integrals. 
This observation 
allows us to cast $H^Q$ in the form  
\begin{equation}\label{quantumHT}
H^Q_{\rho\sigma}(r,q,p) = q^\lambda T_{\lambda\rho\sigma}(q,r,p).
\end{equation}
where $T$ simply represents the rest of the diagram (see Fig.~\ref{Hquantum}).
If we assume now that $T$ has a finite (non-divergent) value in the limit  $q\rightarrow 0$, then 
\begin{equation}\label{limitHQ}
\lim_{q\rightarrow 0}H^Q_{\rho\sigma}(r,q,p) = \lim_{q\rightarrow 0} q^\lambda T_{\lambda\rho\sigma}(q,r,p) = 0.
\end{equation}
So, using \1eq{quantumHT}, the term given in \1eq{badterm} becomes
\begin{equation}\label{badtermT}
{\cal U^{\,\prime}_{\alpha\mu\nu}}(q,r,p) = \frac{q_\alpha q^\lambda}{q^2}F(q^2)m^2(r^2)P_\mu^\rho(r)P_\nu^\sigma(p)T_{\lambda\rho\sigma}(q,r,p) - \binom{r\leftrightarrow p}{\mu \leftrightarrow \nu}.
\end{equation}
When ${\cal U^{\,\prime}_{\alpha\mu\nu}}$ is written in this form, 
it is clear that its pole $1/q^2$ 
is actually compensated by $q_\alpha q^\lambda$,
while the remaining terms cancel themselves when $r=-p$. 
Therefore, the entire term ${\cal U^{\,\prime}_{\alpha\mu\nu}}$ vanishes as $q\rightarrow 0$, 
and should not be included in the ${\cal U}$ part of the $Q^3$ pole vertex.

Finally, returning to the identification $\widetilde{{U}} = \widetilde{{\cal U}}$, 
note that the expression in \1eq{BQQUexpression} does not get modified by the application of 
criterion \n{ii}, because the above argument of discarding $H^Q$ does not apply in this case.
The reason for that is simply the channeling of the momenta in 
$H_{\rho\sigma}(q,r,p)$ and $H_{\rho\sigma}(q,p,r)$ (encoded into the gluon-ghost vertices $\Gamma_\rho$) 
is different from that of $H_{\rho\sigma}(r,q,p)$; 
the momentum entering in the ghost leg is no longer $q$, but rather $r$ or $p$, and the 
corresponding $H^Q$s do not satisfy \1eq{limitHQ}. 
As a result, no terms need be discarded from $\widetilde{{U}}$, 
and, therefore, ${\widetilde{{\cal U}}}^{\,\prime} =0$.

Let us now cast ${\cal U}_{\alpha\mu\nu}(q,r,p)$ and $\widetilde{{\cal U}}_{\alpha\mu\nu}(q,r,p)$
into their canonical form of  \1eq{Ugeneric} and \1eq{backgroundUgeneric}, respectively, namely   
\bea
{\cal U}_{\alpha\mu\nu}(q,r,p) &=& -\frac{q_\alpha}{q^2}I(q^2)B_{\mu\nu}(q,r,p),
\nonumber\\
\widetilde{{\cal U}}_{\alpha\mu\nu}(q,r,p) &=& -\frac{q_\alpha}{q^2}\widetilde{I}(q^2)B_{\mu\nu}(q,r,p); 
\label{canform}
\eea
the minus sign comes from the extra imaginary factor appearing in the Feynman rule of the effective vertex, 
see Fig.~\ref{transitionandB}.
The important point to recognize is that $B_{\mu\nu}(q,r,p)$ is common to both vertices, because 
the legs carrying $r$ and $p$ are quantum ones, and the difference induced due to the 
quantum or background nature of the leg carrying momentum $q$ is entirely encoded into 
the form of the corresponding transition amplitudes, $I(q^2)$ and  $\widetilde{I}(q^2)$, respectively.
Then, \1eq{canform} and \1eq{threegluonBQIU} imply directly the validity of \1eq{BQItransition}.

\section{\label{bqiexp} Diagrammatic demonstration of the BQI}

It is well-known that  the BQIs relating the conventional and BFM Green's functions 
are formally obtained by resorting to the powerful BV formalism. 
On the other hand, the PT (or its generalized version~\cite{Pilaftsis:1996fh}) furnishes an equivalent diagrammatic derivation, 
which makes extensive use of the STIs satisfied by the kernels appearing in the SDEs of the Green's functions 
in question.  
In the previous sections the BQI of \1eq{BQItransition} relating the transition amplitudes  $I(q^2)$ and $\widetilde{I}(q^2)$ 
has been obtained as a self-consistency requirement 
between two SDEs,  
and from the corresponding BQI relating ${\cal U}$ and $\widetilde{{\cal U}}$ 
(sections \ref{bqi} and \ref{excon}, respectively).
The objective of this section is to carry out a PT-guided demonstration of this particular BQI, 
through the systematic conversion of the set of Feynman 
diagrams defining one transition amplitude  
into the set defining the other (shown in Figs.~\ref{transitionandB} and \ref{BFMtransition}). Essentially this conversion proceeds by allowing the 
longitudinal (pinching) momenta contained in the tree-level three-gluon vertex [graph $(\widetilde{d_1})$] to act on the adjacent kernel; therefore, 
as we will see, a crucial ingredient for completing this construction is the knowledge of the STI satisfied by the effective vertex $B_{\mu\nu}$.
In addition, the requirement that the BQI be diagrammatically exact  
imposes a strong constraint on the set of ghost diagrams that 
contribute to the transition amplitude, which may be translated (under mild assumptions) 
into the vanishing of the corresponding subset of $B$ vertices.

\subsection{\label{dem}The PT construction at the level of the transition amplitude}

\begin{figure}[t]
\center{\includegraphics[scale=0.8]{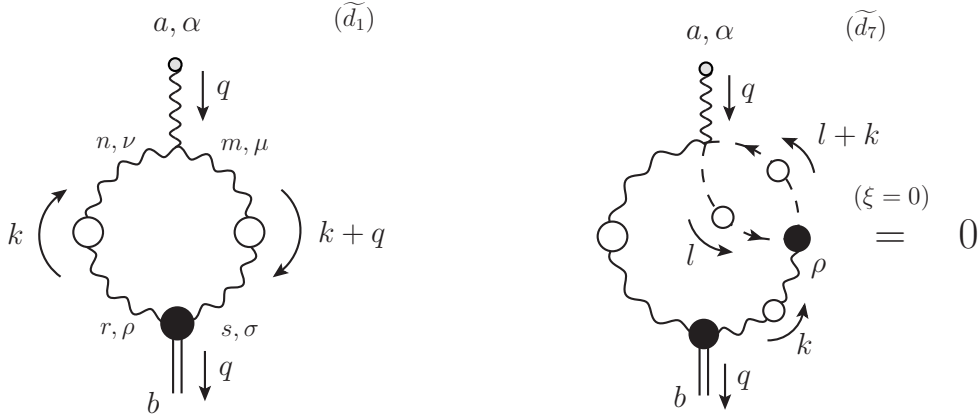}}
\caption{Detailed representation of diagrams $(\widetilde{d_1})$ and $(\widetilde{d_7})$. 
Diagram $(\widetilde{d_1})$ has a symmetry factor $1/2$; 
diagrams $(\widetilde{d_6})$ and $(\widetilde{d_7})$ vanish in the Landau gauge.}
\label{one-looptransition}
\end{figure}

Let us start by considering the diagram $(\widetilde{d_1})$ of the background transition amplitude, shown in 
Fig.~\ref{one-looptransition}, whose contribution is given by
\begin{equation}\label{d1noLandau}
(\widetilde{d_1})_\alpha = \frac{i}{2}C_A\int_k \widetilde{\Gamma}^{(0)}_{\alpha\mu\nu}\Delta^{\mu\sigma}(k+q)\Delta^{\nu\rho}(k)B_{\rho\sigma}.
\end{equation}
In order to avoid notational clutter we will suppress the arguments of the momenta in 
the vertices, which can be easily recovered from the figures. In what follows, the coupling and color dependence of the diagrams comprising the evaluation of $\widetilde{I}_\alpha(q)$ is restored by setting $(\widetilde{d_i})^{ab}_\alpha  = g\delta^{ab}(\widetilde{d_i})_\alpha$.

We know that the tree-level vertex $\widetilde{\Gamma}^{(0)}_{\alpha\mu\nu}$ contains 
terms which are proportional to $\xi^{-1}$, so that one cannot take directly the limit 
$\xi=0$ to work in the Landau gauge. Specifically, the tree-level part of the $BQ^2$ vertex 
is given by
\begin{equation}\label{decompositionvertex}
\widetilde{\Gamma}^{(0)}_{\alpha\mu\nu}(q,-k-q,k)=\Gamma^{(0)}_{\alpha\mu\nu}(q,-k-q,k)-\frac{1}{\xi}\Gamma^{P}_{\alpha\mu\nu}(q,-k-q,k),
\end{equation}
where the purely longitudinal ``pinch part'' $\Gamma^{P}$ reads
\begin{equation}\label{pinchpart}
\Gamma^{P}_{\alpha\mu\nu}(q,-k-q,k)=g_{\alpha\mu}k_\nu + g_{\alpha\nu}(k+q)_\mu,
\end{equation}
and
\begin{equation}\label{tree-levelQQQ}
\Gamma^{(0)}_{\alpha\mu\nu}(q,-k-q,k) = -(2k+q)_\alpha g_{\mu\nu} + (k-q)_\mu g_{\alpha\nu} + (2q+k)_\nu g_{\alpha\mu}
\end{equation}
is the tree-level value of the conventional three-gluon vertex. Using then the 
decomposition \1eq{decompositionvertex}, we obtain from \1eq{d1noLandau}
\begin{equation}\label{d1pinch}
(\widetilde{d_1})_\alpha = (d_1)_\alpha - \frac{i}{2\xi}C_A\int_k \Gamma^{P}_{\alpha\mu\nu}\Delta^{\mu\sigma}(k+q)\Delta^{\nu\rho}(k)B_{\rho\sigma}.
\end{equation}
Now, after the shifts $k+q\mapsto k$ and $k\mapsto -k$, and applying the antisymmetry 
property of the effective vertex, namely $B_{\rho\sigma} = -B_{\sigma\rho}$, the two 
contributions coming from the pinch part of the vertex in \1eq{d1pinch} sum up and cancel 
the symmetry factor, giving the result
\begin{equation}\label{d1sum}
(\widetilde{d_1})_\alpha = (d_1)_\alpha - \frac{i}{\xi}C_A\int_k k_\nu\Delta^{\nu\rho}(k)\Delta^\sigma_\alpha(k+q)B_{\rho\sigma}.
\end{equation}
Therefore, employing the identity
\begin{equation}\label{propagatorcontraction}
\frac{1}{\xi}k_\nu\Delta^{\nu\rho}(k)=\frac{k^\rho}{k^2},
\end{equation}
the $\xi^{-1}$ term cancels, and we can project \1eq{d1sum} to the Landau gauge ($\xi=0$),
\begin{equation}\label{d1Landau}
(\widetilde{d_1})_\alpha = (d_1)_\alpha - iC_A\int_k\frac{k^\rho}{k^2}\Delta_\alpha^\sigma(k+q)B_{\rho\sigma},
\end{equation}
where the gluon propagator assumes a totally transverse form, \ie $\Delta_{\mu\nu}(k) = \Delta(k^2)P_{\mu\nu}(k)$.

At this point, in order to evaluate the integral on the rhs of \1eq{d1Landau}, the knowledge 
of the STI satisfied by the effective vertex $B_{\rho\sigma}$ is required. 
Observe then that, if we compare \1eq{BQQUexpression} with \1eq{canform}, the following structure can 
be identified as the $B_{\mu\nu}$,
\begin{eqnarray}\label{effectiveB}
\widetilde{I}(q^2)B_{\mu\nu}(q,r,p) &=& -m^2(r^2)P_\mu^\rho(r)P_{\rho\nu}(p) 
+\frac{r_\mu}{2}D(r^2)\widetilde{m}^2(q^2)[g_\nu^\rho + P_\nu^\rho(p)] \Gamma_\rho(p,q,r) 
\nonumber\\
&-& \binom{r\leftrightarrow p}{\mu \leftrightarrow \nu}.
\end{eqnarray} 
Therefore, contracting \1eq{effectiveB} with respect to the momentum $r^\mu$, and breaking  
the transverse projector $P_\nu^\rho(p)$, we find
\begin{eqnarray}\label{almostSTIB}
\widetilde{I}(q^2)r^\mu B_{\mu\nu}(q,r,p) &=& F(r^2)\widetilde{m}^2(q^2)\Gamma_\nu(p,q,r) \nonumber \\
&-& \frac{p_\nu}{2p^2}\widetilde{m}^2(q^2)[F(r^2)p^\rho\Gamma_\rho(p,q,r) + F(p^2)r^\rho\Gamma_\rho(r,q,p)].
\end{eqnarray}
Now, the second term on the rhs does not contribute, since  $p_\nu \rightarrow (k+q)_{\sigma}$ 
is annihilated by the transverse projector of $\Delta_\alpha^\sigma(k+q)$ in \1eq{d1Landau}.

Thus, employing the STI \1eq{almostSTIB}, adapted to our kinematical configuration, 
\ie \mbox{$r=-k$} and $p=k+q$, together with \1eq{QBmassformula}, 
\1eq{d1Landau} becomes
\begin{eqnarray}\label{d1Lambda}
(\widetilde{d_1})_\alpha &=& (d_1)_\alpha + ig^2C_A I(q^2)\int_k \Delta_\alpha^\sigma(k+q)D(k)\Gamma_\sigma \nonumber \\
&=& (d_1)_\alpha - ig^2C_A I(q^2)q^\lambda \int_k \Delta_\alpha^\sigma(k+q)D(k)H_{\lambda\sigma},
\end{eqnarray}
where, in the second line, \1eq{ghH} has been used.

At this point, one recognizes on the rhs of \1eq{d1Lambda}
the appearance of the auxiliary two-point function $\Lambda$, defined in \1eq{Lambda}; so, we obtain 
\bea
\label{almostBQItransition}
(\widetilde{d_1})_\alpha &=& (d_1)_\alpha + I(q^2)q^\lambda \Lambda_{\alpha\lambda}(q) 
\nonumber \\
&=& (d_1)_\alpha + [G(q^2) + L(q^2)]I_\alpha(q).
\eea
Finally, since $L(0)=0$, one can drop the term $L(q^2)$, exactly as was done in the previous section (and with the same justification),  
finally arriving at
\begin{equation}\label{incompleteBQItransition}
(\widetilde{d_1})_\alpha = (d_1)_\alpha + G(q^2)I_\alpha(q).
\end{equation}

Let us next observe that {\it (i)} diagrams $(d_3)$ and $(d_4)$ in Fig.~\ref{transitionandB} can be converted 
automatically to background ones, given that the tree-level 
vertex $BQ^3$ is identical to the conventional four gluon vertex $Q^4$, and that {\it (ii)}
diagrams $(\widetilde{d_6})$ and $(\widetilde{d_7})$ vanish in the Landau gauge, see Fig.~\ref{one-looptransition}; 
this happens because, after integration, the contribution of the ghost loop nested inside them 
will be proportional to $k^{\rho}$, and it will vanish when contracted with the $\Delta_{\rho\rho'}(k)$.
So,
\be
(\widetilde{d_3}) = (d_3)\,, \,\,\,\, (\widetilde{d_4}) = (d_4)\,, \,\,\,\, (\widetilde{d_6})=(\widetilde{d_7}) =0\,.
\label{residual}
\ee
Therefore, comparing \1eq{incompleteBQItransition} and \1eq{residual} with 
\1eq{BQItransition}, we see that, in order for the full BQI to be diagrammatically realized, the relation
\begin{equation}\label{constraintghost}
(d_2) = (\widetilde{d_2})+(\widetilde{d_5})+(\widetilde{d_8})
\end{equation}
should be satisfied. 

\subsection{\label{ghostcon} A constraint on the ghost sector}

In the previous subsection we have derived \1eq{constraintghost} from the requirement that the basic BQI \1eq{BQItransition} be diagrammatically realized.
The common characteristic of all diagrams comprising \1eq{constraintghost} is that they contain vertices $B$ which  
mix the nonperturbative massless bound-state with two ghosts.
Given that the nature of these vertices is practically unknown, 
it would be interesting to inquire what restriction 
\1eq{constraintghost} may impose on their structure, and whether this restriction is compatible with other pieces of nonperturbative information 
on the ghost sector of Yang-Mills theories.

In the formulas that follow, color factors are shown explicitly, while the arguments of the momenta in 
the vertices are suppressed as before.
 Then, the diagrams appearing in \1eq{constraintghost} are given by
\bea
(d_2)^{ab}_\alpha &=& -igC_A\delta^{ab}\int_k \Gamma^{(0)}_\alpha D(k)D(k+q)B\,,
\nonumber\\
(\widetilde{d}_2)^{ab}_\alpha &=& -igC_A\delta^{ab}\int_k \widetilde{\Gamma}^{(0)}_\alpha D(k)D(k+q)B\,, 
\nonumber\\
(\widetilde{d}_5)^{ab}_\alpha &=& g^3C_A^2\delta^{ab}\int_k \widetilde{\Gamma}^{(0)}_{\alpha\mu}D(k)D(k+q)B\int_l D(k+l)\Delta^{\mu\nu}(l)\Gamma_\nu\,,
\nonumber\\
(\widetilde{d}_8)^{ab}_\alpha &=& g^2f^{adx}f^{xcm}\int_k\int_l D(k+q)D(k+l)\Delta_\alpha^\nu(l)B_\nu^{bcmd}\,.
\label{BFMd2ghost}
\eea
Observe that the pole part of the 
internal gluon-ghost vertex $\Gamma_\nu$ does not contribute to $(\widetilde{d}_5)$ in the Landau gauge, 
due to the longitudinality condition that it ought to satisfy. Therefore, only the regular part of this vertex survives in $(\widetilde{d}_5)$.

\begin{figure}[t]
\center{\includegraphics[scale=0.65]{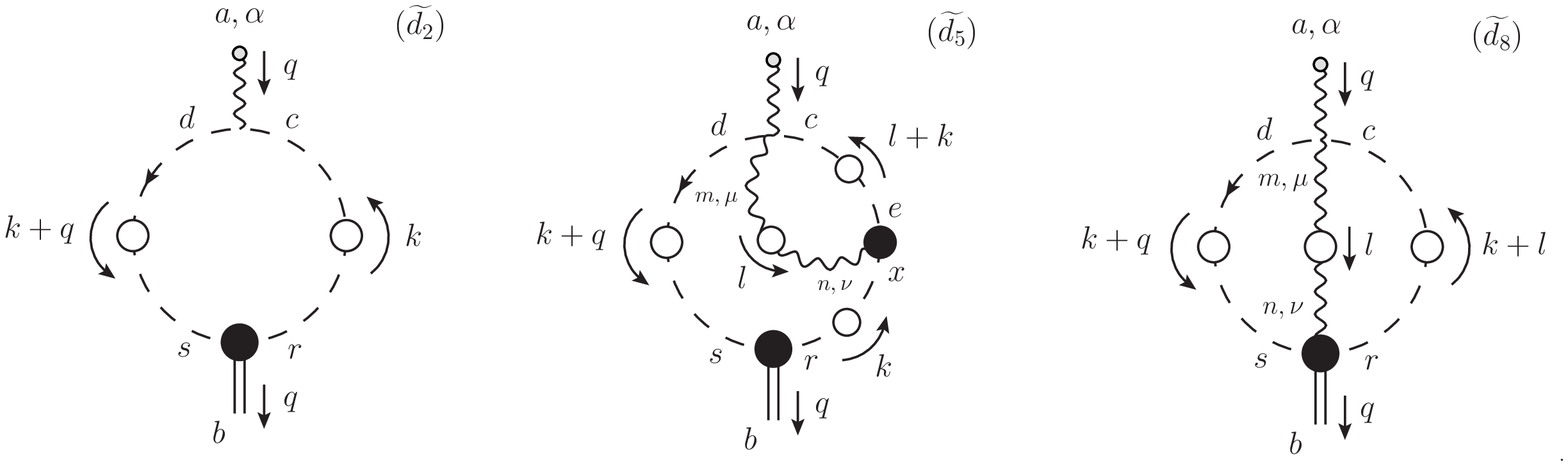}}
\caption{Detailed representation of the diagrams $(\widetilde{d_2})$, $(\widetilde{d_5})$, and $(\widetilde{d_8})$, appearing in \1eq{constraintghost}.}
\label{ghostloops}
\end{figure}

We will next use the SDE satisfied by the ghost propagator $D(k)$, written as~\cite{Binosi:2008qk}
\begin{equation}\label{SDEghostpropagator}
g^2C_A D(k)\int_l D(k+l)\Delta^{\mu\nu}(l)\Gamma_\nu = ik_\alpha[D(k)-D^{(0)}(k)],
\end{equation}
to cast $(\widetilde{d}_5)$ in the form
\begin{equation}\label{compactd5}
(\widetilde{d}_5)^{ab}_\alpha = igC_A\delta^{ab}\int_k k_\alpha[D(k)-D^{(0)}(k)]D(k+q)B.
\end{equation}

Substituting the above results into \1eq{constraintghost}, we arrive at the integral constraint 
\begin{equation}\label{integralconstraint}
iC_A\delta^{ab}\int_k \frac{k_\alpha}{k^2}D(k+q)B = gf^{adx}f^{xcm}\int_k\int_l D(k+q)D(k+l)\Delta_\alpha^\nu(l)B_\nu^{bcmd}\,,
\end{equation}
relating $B$ and $B_\nu$.

We next turn to the full BFM gluon-ghost vertex, $\widetilde{\Gamma}$, which satisfies the following all-order WI~\cite{Aguilar:2006gr,Binosi:2008qk} 
\begin{equation}\label{BFMgluon-ghostWI}
q^\alpha\widetilde{\Gamma}_\alpha(q,r,p) = D^{-1}(r^2) - D^{-1}(p^2).
\end{equation}

Let us suppose that, due to nonperturbative dynamics,  $\widetilde{\Gamma}$ acquires a pole part; then, 
in full analogy to \1eq{modifiedvertex}, we define  
\begin{equation}\label{fullBFMgluon-ghost}
\widetilde{\Gamma}'_\alpha(q,r,p) = \widetilde{\Gamma}_\alpha(q,r,p) + \widetilde{V}_\alpha(q,r,p),
\end{equation}
as the sum of a regular part and a pole part. In addition, and again in analogy with \1eq{massive}, we will assume that also 
the ghost propagator acquires a mass term, to be denoted by $m^2_{c}(q^2)$, and so 
\be
D^{-1}(q^2) = q^2 F^{-1}(q^2) \longmapsto D_m^{-1}(q^2)= q^2 F^{-1}_m (q^2)- m^2_{c}(q^2).
\label{massiveghost}
\ee
Then, it is clear that we must have   
\bea
q^\alpha\widetilde{\Gamma}_\alpha(q,r,p) &=& r^2 F^{-1}_m (r^2) - p^2 F^{-1}_m (p^2)\,,
\nonumber\\
q^\alpha \widetilde{V}_\alpha(q,r,p) &=& m^2_{c}(p^2) - m^2_{c}(r^2)\,,
\eea
so that finally 
\be
q^\alpha \widetilde{\Gamma}'_\alpha(q,r,p) =  D_m^{-1}(r^2) - D_m^{-1}(p^2)\,,
\ee
as it should. 

If we now assume the existence of a massless pole in the $q$-channel which, in addition, satisfies the longitudinality condition
\begin{equation}\label{totallylongV}
P^\beta_\alpha(q)\widetilde{V}_\beta(q,r,p) = 0,
\end{equation}
then, the WI obeyed by $\widetilde{V}_\alpha(q,r,p)$, may be solved, yielding
\begin{equation}\label{poleBFMgluon-ghost}
\widetilde{V}_\alpha(q,r,p) = \frac{q_\alpha}{q^2}[m^2_{c}(p^2) - m^2_{c}(r^2)].
\end{equation}
Note that $\widetilde{V}_\alpha(q,r,p)$ is purely of the $\widetilde{{\cal U}}$ type (no $\widetilde{{\cal R}}$ terms).

Then, using the corresponding expression \1eq{backgroundUgeneric} for this vertex
\begin{equation}\label{Vgluon-ghostpole}
\widetilde{V}_\alpha(q,r,p) = -\frac{q_\alpha}{q^2}\widetilde{I}(q^2)B(q,r,p),
\end{equation}
and equating with \1eq{poleBFMgluon-ghost}, we obtain a relation between the effective vertex $B$ 
and the mass-like term in the ghost propagator
\begin{equation}\label{relationBf}
\widetilde{I}(q^2)B(q,r,p) = m^2_{c}(r^2) - m^2_{c}(p^2).
\end{equation}
Now, large-volume lattice 
results~\cite{Cucchieri:2007md,Cucchieri:2007rg,Cucchieri:2009zt,Bogolubsky:2007ud,Bogolubsky:2009dc}, 
together with a variety of analytic studies~\cite{RodriguezQuintero:2010ss,Dudal:2008sp,Boucaud:2008ky}, 
establish that {\it (i)}  
the ghost propagator in the Landau gauge remains massless, which means that $m^2_{c}(q^2)=0$. 
{\it (ii)} The dressing function $F_m(q^2)$ is infrared-finite; this happens essentially due to the fact that the 
gluon mass $m^2(q^2)$ regulates it, in the same way as $J_m$, namely through the qualitative replacement 
$\ln q^2 \to \ln(q^2+m^2)$ [see discussion after \1eq{massive}].
Therefore, since $m^2_{c}(q^2)=0$ but $\widetilde{I}(q^2) \neq 0$, \1eq{relationBf} implies that $B=0$. 

Returning to the constraint of \1eq{integralconstraint} and setting $B=0$, we have that the integral on the rhs 
involving $B_\nu$ must vanish also.  Even though from this statement one cannot mathematically conclude that 
$B_\nu$ must vanish identically, this seems to be the most reasonable scenario. Indeed, the 
rather  fine-tuned alternative would invoke  
a subtle conspiracy between the various form factors comprising $B_\nu$, which ought to 
combine in such a way (and change signs correspondingly) as to make this integral vanish.

\section{\label{SDEvsBSE} Comparison of the two mass\,-generating formalisms}

In  this section  we  first derive the  full  expression for  the
transition amplitude in  the Landau gauge, and then   
we show that, when judiciously combined with 
\1eq{QQmassformula} and \1eq{squareI}, it 
gives rise to an integral equation for the gluon mass, which {\it coincides exactly} 
with the one obtained within the SDE formalism of~\cite{Binosi:2012sj}.
We then proceed to a brief study of how the distinct approximations employed within the two 
formalism lead to differences in the form of the gluon masses obtained.

\subsection{\label{self} Exact formal equivalence}

\begin{figure}[ht]
\center{\includegraphics[scale=0.7]{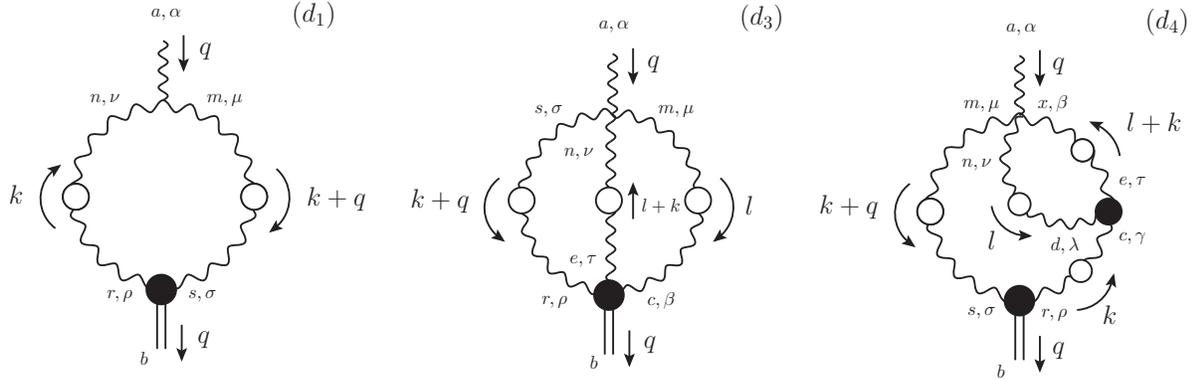}}
\caption{Configuration for the diagrams of $I_{\alpha}(q)$ that contain only gluon loops. 
The symmetry factors for the diagrams are $S(d_1,d_3,d_4)=(1/2,1/6,1/2)$.}
\label{gluonloops}
\end{figure}

From the analysis presented in the previous section we have concluded that 
diagrams with ghost loops do not contribute to the transition amplitude. Furthermore, one may demonstrate 
that diagram $(d_3)$ in Fig.~\ref{gluonloops}, which contains the effective vertex $B_{\rho\tau\beta}$ 
mixing three quantum gluons with the bound state, vanishes in the Landau gauge. Effectively, 
with the configuration shown in Fig.~\ref{gluonloops}, we can write down the contribution of
this diagram,
\begin{equation}\label{d3color}
(d_3)_\alpha^{ab} = \frac{1}{6}\int_k\int_l\Gamma^{(0)amns}_{\alpha\mu\nu\sigma}\Delta^{\sigma\rho}_{sr}(k+q)\Delta^{\nu\tau}_{ne}(l+k)\Delta^{\mu\beta}_{mc}(l)B_{\rho\tau\beta}^{brec}.
\end{equation}
As in the preceding sections, the arguments of the momenta in the vertices have been suppressed. Observe that, in order to evaluate this diagram in the Landau gauge, one should know how three transverse projectors act over the effective vertex $B_{\rho\tau\beta}$. Consider then the decomposition \1eq{Valtsep} for the $BQ^3$ pole vertex,
\begin{equation}\label{BQQQdecomposition}
\widetilde{V}_{\lambda\rho\tau\beta}(q,r,p,l) = \widetilde{{\cal U}}_{\lambda\rho\tau\beta}(q,r,p,l) + \widetilde{{\cal R}}_{\lambda\rho\tau\beta}(q,r,p,l),
\end{equation}
where, according with the discussion carry out in Section~\ref{massde}, the $\widetilde{{\cal R}}$ part of the vertex satisfies the transversality condition
\begin{equation}\label{RpartBQQQ}
P^{\sigma\rho}(r)P^{\nu\tau}(p)P^{\mu\beta}(l)\widetilde{{\cal R}}_{\lambda\rho\tau\beta}(q,r,p,l) = 0.
\end{equation}
Thus, when \1eq{totlon} is applied for the case of the $BQ^3$ pole vertex, namely, four transverse projectors canceling the vertex, we obtain from \1eq{BQQQdecomposition} the following result
\begin{equation}\label{relationUV}
P^{\sigma\rho}(r)P^{\nu\tau}(p)P^{\mu\beta}(l)\widetilde{{\cal U}}_{\alpha\rho\tau\beta}(q,r,p,l) = \frac{q_\alpha}{q^2}P^{\sigma\rho}(r)P^{\nu\tau}(p)P^{\mu\beta}(l)q^\lambda\widetilde{V}_{\lambda\rho\tau\beta}(q,r,p,l).
\end{equation}
On the other hand \1eq{backgroundUgeneric} becomes for the $\widetilde{{\cal U}}$ part of this pole vertex,
\begin{equation}\label{UpartBQQQ}
\widetilde{{\cal U}}_{\alpha\rho\tau\beta}(q,r,p,l)=ig\frac{q_\alpha}{q^2}\widetilde{I}(q^2)B_{\rho\tau\beta}(q,r,p,l).
\end{equation}
Therefore, applying three transverse projectors on \1eq{UpartBQQQ} and equating the result with \1eq{relationUV} 
we can relate the effective vertex $B_{\rho\tau\beta}$ with the WI satisfied by the $BQ^3$ pole vertex when it is contracted with respect to the momentum of the background gluon leg, \ie
\begin{equation}\label{relationBV}
ig\widetilde{I}(q^2)P^{\sigma\rho}(r)P^{\nu\tau}(p)P^{\mu\beta}(l)B_{\rho\tau\beta}(q,r,p,l) = P^{\sigma\rho}(r)P^{\nu\tau}(p)P^{\mu\beta}(l)q^\lambda\widetilde{V}_{\lambda\rho\tau\beta}(q,r,p,l).
\end{equation}
Once the above connection has been established through \1eq{relationBV}, 
one may repeat the steps presented in subsection VI B of~\cite{Binosi:2012sj}, thus demonstrating the vanishing of diagram $(d_3)$.

Therefore, the full transition amplitude will be given in the Landau gauge solely by the sum of diagrams $(d_1)$ and $(d_4)$. 
Thus, applying the conventions of Fig.~\ref{gluonloops}, we obtain for diagram $(d_1)$ the expression,
\begin{equation}\label{d1gluons}
(d_1)_\alpha = \frac{i}{2}C_A \int_k \Gamma_{\alpha\mu\nu}^{(0)} \Delta^{\mu\sigma}(k+q) \Delta^{\nu\rho}(k) B_{\rho\sigma}.
\end{equation}
One observes that the above integral has one free Lorentz index, which only can be saturated by the external momentum $q$. 
So, using the elementary WI for the tree-level three-gluon vertex,
\begin{equation}\label{elementaryWI}
q^\lambda \Gamma^{(0)}_{\lambda\mu\nu}(q,-k-q,k) = k^2 P_{\mu\nu}(k) - (k+q)^2 P_{\mu\nu}(k+q),
\end{equation}
and after the appropriate shifts in the integrated momentum, we deduce in the Landau gauge the result,
\begin{equation}\label{finald1}
(d_1)_\alpha = \frac{q_\alpha}{q^2}q^\lambda (d_1)_\lambda = iC_A \frac{q_\alpha}{q^2}\int_k k^2 \Delta_\mu^\rho(k)\Delta^{\mu\sigma}(k+q) B_{\rho\sigma}.
\end{equation}

Now, diagram $(d_4)$ contains the tree-level four-gluon vertex $\Gamma^{(0)}_{\alpha\mu\nu\beta}$. 
After the color algebra and using the standard Feynman rule for this vertex, we obtain for the prefactor of the diagram,
\begin{equation}\label{prefactord4}
\frac{i}{2}gf^{bcm}f^{cxn}\Gamma^{(0)amnx}_{\alpha\mu\nu\beta}=\frac{3}{4}g^3C_A^2(g_{\alpha\nu}g_{\mu\beta}-g_{\alpha\beta}g_{\mu\nu})\delta^{ab}.
\end{equation}
Thus, we get in the Landau gauge the following expression,
\begin{equation}\label{d4gluons}
(d_4)_\alpha = \frac{3}{4}g^2C_A^2(g_{\alpha\nu}g_{\mu\beta}-g_{\alpha\beta}g_{\mu\nu})\int_k \Delta^{\mu\sigma}(k+q)\Delta^{\rho\gamma}(k)Y_\gamma^{\nu\beta}(k)B_{\rho\sigma},
\end{equation}
where we have defined the loop integral
\begin{equation}\label{Yintegral}
Y_\gamma^{\nu\beta}(k) = \int_l \Delta^{\nu\lambda}(l) \Delta^{\beta\tau}(k+l)\Gamma_{\gamma\tau\lambda}.
\end{equation}
Note that in the Landau gauge, the fully-dressed three-gluon vertex appearing in this integral only 
contains the regular part, since the transverse projectors of the gluon propagators trigger \1eq{totlon} for that vertex.

As before, the integral in \1eq{d4gluons} only can be saturated by the external momentum $q$. 
Moreover, due to the Bose symmetry of the three-gluon vertex is straightforward to show 
that the integral \1eq{Yintegral} is antisymmetric under the $\nu\leftrightarrow\beta$ exchange, 
and given also the antisymmetry of the prefactor under the same exchange, one can write
\begin{equation}\label{Yformfactor}
Y_\gamma^{\nu\beta}(k)=(k^\nu g_\gamma^\beta - k^\beta g_\gamma^\nu)Y(k^2) \quad ; \quad Y(k^2)=\frac{1}{d-1}\frac{k_\nu}{k^2}g_\beta^\gamma Y_\gamma^{\nu\beta}(k).
\end{equation}
With these observations, \1eq{d4gluons} can be cast in the form
\begin{equation}\label{finald4}
(d_4)_\alpha = \frac{q_\alpha}{q^2}q^\lambda (d_4)_\lambda = \frac{3}{2}g^2C_A^2\frac{q_\alpha}{q^2}\int_k[(kq)g_{\mu\gamma} + q_\mu q_\gamma]Y(k^2)\Delta^{\mu\sigma}(k+q)\Delta^{\rho\gamma}(k)B_{\rho\sigma}.
\end{equation}

\begin{figure}[t]
\center{\includegraphics[scale=0.7]{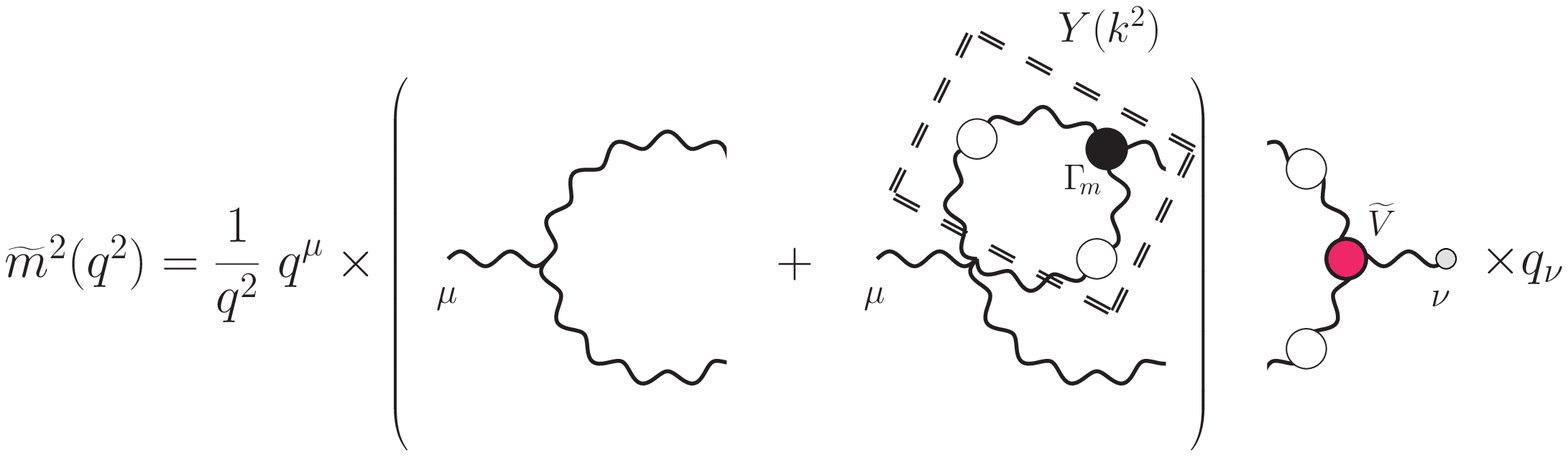}}
\caption{Concise diagrammatic representation of the combined operations 
leading to the all-order gluon mass equation; ${\widetilde m}^2 (q^2)$ is defined in \1eq{BQImasses}.}
\label{diagrammaticmass}
\end{figure}

Employing now the results obtained for diagrams $(d_1)$ and $(d_4)$, 
we derive the complete expression of the transition amplitude in the Landau gauge [see \1eq{scalarcofactor}],
\begin{eqnarray}\label{completefulltransition}
I(q^2) &=& \frac{q^\alpha}{q^2} [(d_1) + (d_4)]_\alpha = \frac{i}{q^2}C_A\int_k k^2\Delta_\mu^\rho(k)\Delta^{\mu\sigma}(k+q)B_{\rho\sigma} \nonumber \\
&+& \frac{3}{2}\frac{g^2C_A^2}{q^2}\int_k[(kq)g_{\mu\gamma} + q_\mu q_\gamma]Y(k^2)\Delta^{\mu\sigma}(k+q)\Delta^{\rho\gamma}(k)B_{\rho\sigma}.
\end{eqnarray}
To proceed further, we decompose the effective vertex $B_{\rho\sigma}$ in the tensor basis 
\begin{equation}\label{LorentzB}
B_{\rho\sigma} = B_1 g_{\rho\sigma} + B_2 q_\rho q_\sigma + B_3 (k+q)_\rho (k+q)_\sigma + B_4 k_\rho q_\sigma + B_5 k_\rho (k+q)_\sigma.
\end{equation}
When this decomposition is inserted in \1eq{completefulltransition}, only the form factor $B_1$ survives, since in~\cite{Aguilar:2011xe} was shown that $B_2=0$ and the rest of form factors are canceled in the Landau gauge by the transverse projectors. Therefore, \1eq{completefulltransition} becomes
\begin{eqnarray}\label{transitionB1general}
I(q^2) &=& \frac{i}{q^2}C_A\int_k k^2\Delta_\mu^\rho(k)\Delta^\mu_\rho(k+q)B_1 \nonumber \\
&+& \frac{3}{2}\frac{g^2C_A^2}{q^2}\int_k[(kq)g_{\mu\gamma} + q_\mu q_\gamma]Y(k^2)\Delta^\mu_\rho(k+q)\Delta^{\rho\gamma}(k)B_1,
\end{eqnarray}
which, quite remarkably, 
allows to express the full transition amplitude, for general value of $q^2$, in terms 
of one single form factor, namely, $B_1$. 
At this point we arrive at the crucial observation which enables us 
to relate the mass equation obtained in the context of the 
PT-BFM formalism with the bound-state formalism showing that, indeed both formalisms are self-consistent and interconnected. 
From \1eq{effectiveB} one may obtain, after identify the $g_{\mu\nu}$ component, a closed expression for the form factor $B_1$ 
in terms of the gluon mass
\bea
\widetilde{I}(q^2)B_1(q,r,p) &=& [1+ G(q^2)]I(q^2)B_1(q,r,p) 
\nonumber \\
&=& m^2(p^2) - m^2(r^2),
\label{relB1mass}
\eea
where the BQI \1eq{BQItransition} has been applied. Therefore, using this relation and after a straightforward rearrangement \1eq{transitionB1general} yields
\begin{equation}\label{squareI}
I^2(q^2) = \frac{iC_A}{1+G(q^2)}\frac{1}{q^2}\int_k \Delta_{\gamma\rho}(k)\Delta_\mu^\rho(k+q){\cal K}_{\s{SD}}^{\gamma\mu}(q,k)m^2(k^2).
\end{equation}
In this expression, the quantity 
\begin{eqnarray}\label{SDKernel}
{\cal K}_{\s{SD}}^{\gamma\mu}(q,k) &=& g^{\gamma\mu}[(k+q)^2 - k^2]\bigg\lbrace 1 + \frac{3}{4}ig^2C_A[Y(k+q)+Y(k)]\bigg\rbrace \nonumber \\
&+& \frac{3}{4}ig^2C_A(q^2g^{\gamma\mu} - 2q^\gamma q^\mu)[Y(k+q) - Y(k)].
\end{eqnarray}
represents the $SD$ kernel obtained from the sum of the two graphs on the rhs of the equation depicted in Fig.~\ref{diagrammaticmass}. 
Thus, using the mass formula \1eq{QQmassformula} on the lhs of \1eq{squareI}, we arrive to the result
\begin{equation}\label{masseq}
m^2(q^2) = \frac{ig^2C_A}{1+G(q^2)}\frac{1}{q^2}\int_k m^2(k^2) \Delta_{\gamma\rho}(k)\Delta_\mu^\rho(k+q){\cal K}_{\s{SD}}^{\gamma\mu}(q,k)\,.
\end{equation}
Quite remarkably, this result  coincides {\it exactly}  with the  full mass
equation  derived in~\cite{Binosi:2012sj}  from the  SDE of  the gluon
propagator, following  the  procedure   shown pictorially in
Fig.~\ref{diagrammaticmass}, whose formal aspects are 
considerably  different compared to those of the massless bound-state formalism. 
The emergence of an identical  dynamical  equation for  the gluon  mass 
provides an  impressive self-consistency check between these
two formalisms.

\subsection{\label{VS} Differences in the numerical implementation}

At this point 
it is important to recognize that, although formally equivalent, 
the two approaches [``SDE'' vs ``massless bound-state''] entail vastly different procedures  
for obtaining the desired quantity, namely the functional form of $m^2 (q^2)$. 
Given that in practice approximations must be carried out to 
the fundamental equations of both formalisms, the results obtained for 
$m^2 (q^2)$ will be in general different; indeed, the formal coincidence 
proved above is only valid when all equations involved are treated exactly.

\begin{figure}[ht]
\center{\includegraphics[scale=0.65]{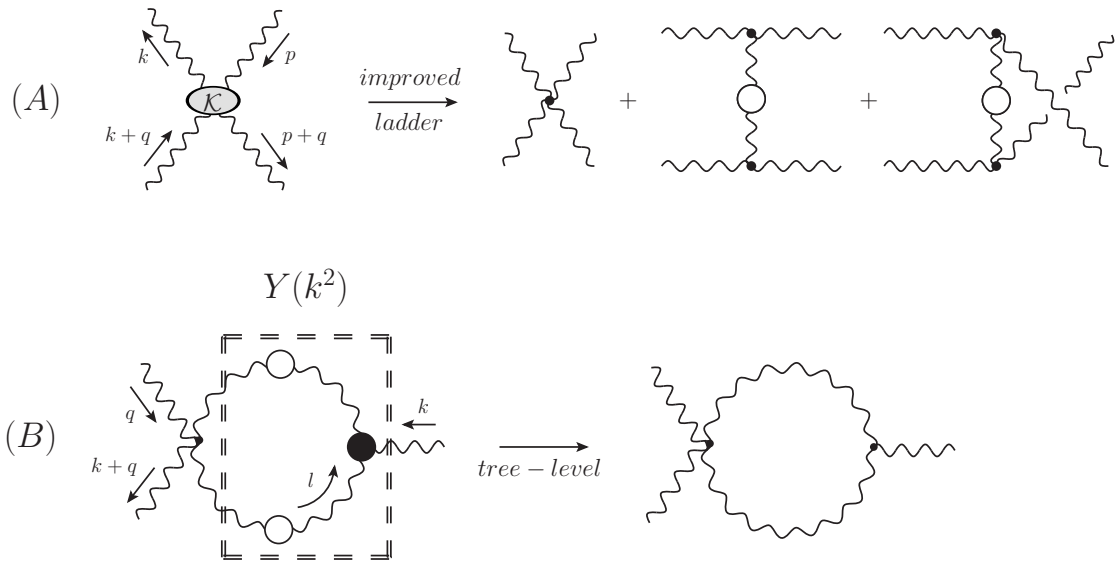}}
\caption{
(A) The kernel ${\cal K}$ of the BSE satisfied by $B_1'$ and the approximation introduced in~\cite{Aguilar:2011xe}.
(B) The full kernel ${\cal K}_{\s{SD}}$ of the mass equation, and the approximation employed in~\cite{Binosi:2012sj}.}
\label{SDEvsBS}
\end{figure}

\begin{figure}[!t]
\begin{minipage}[b]{0.45\linewidth}
\includegraphics[scale=0.54]{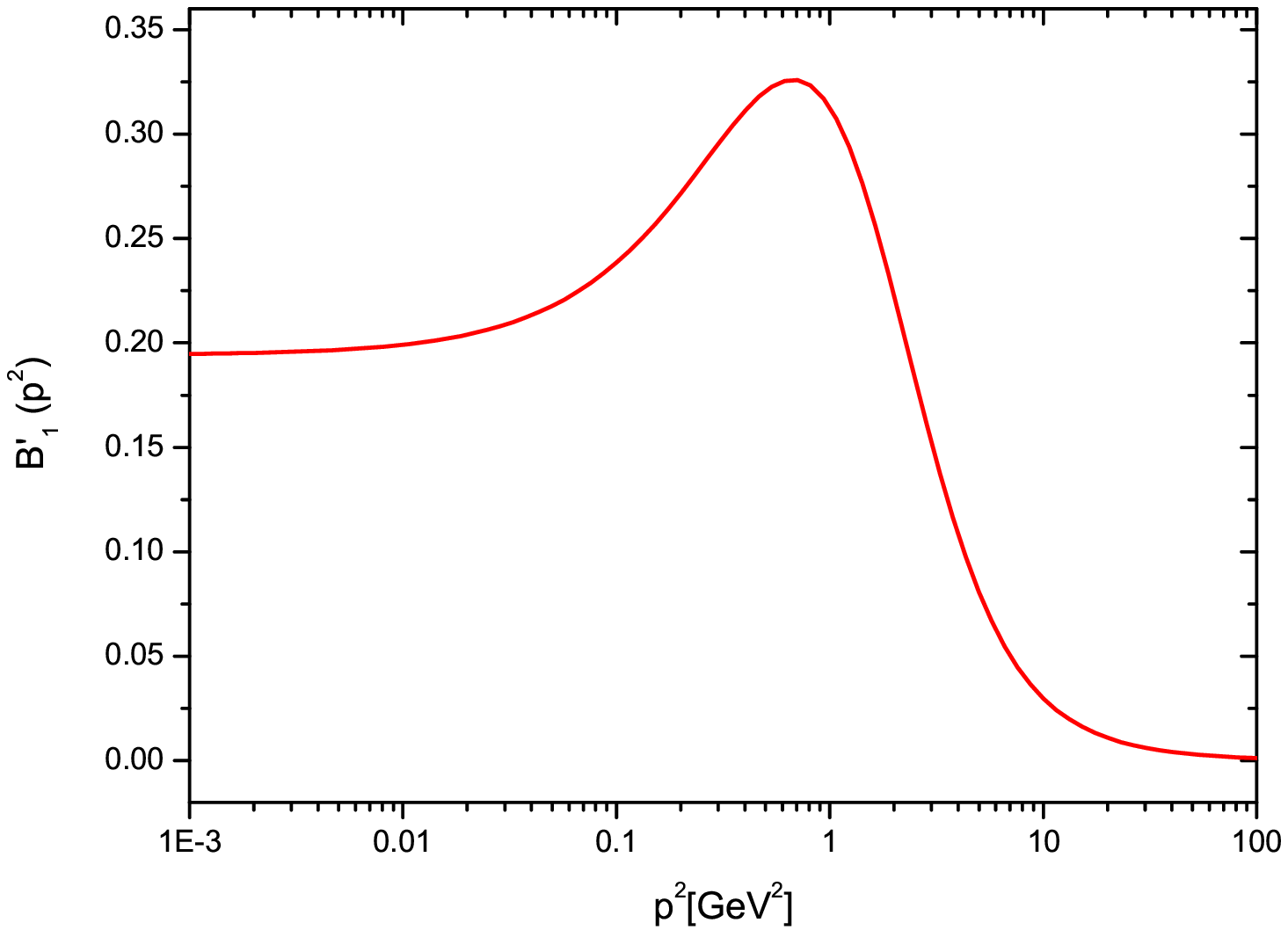}
\end{minipage}
\hspace{0.58cm}
\begin{minipage}[b]{0.50\linewidth}
\hspace{-1cm}
\includegraphics[scale=0.53]{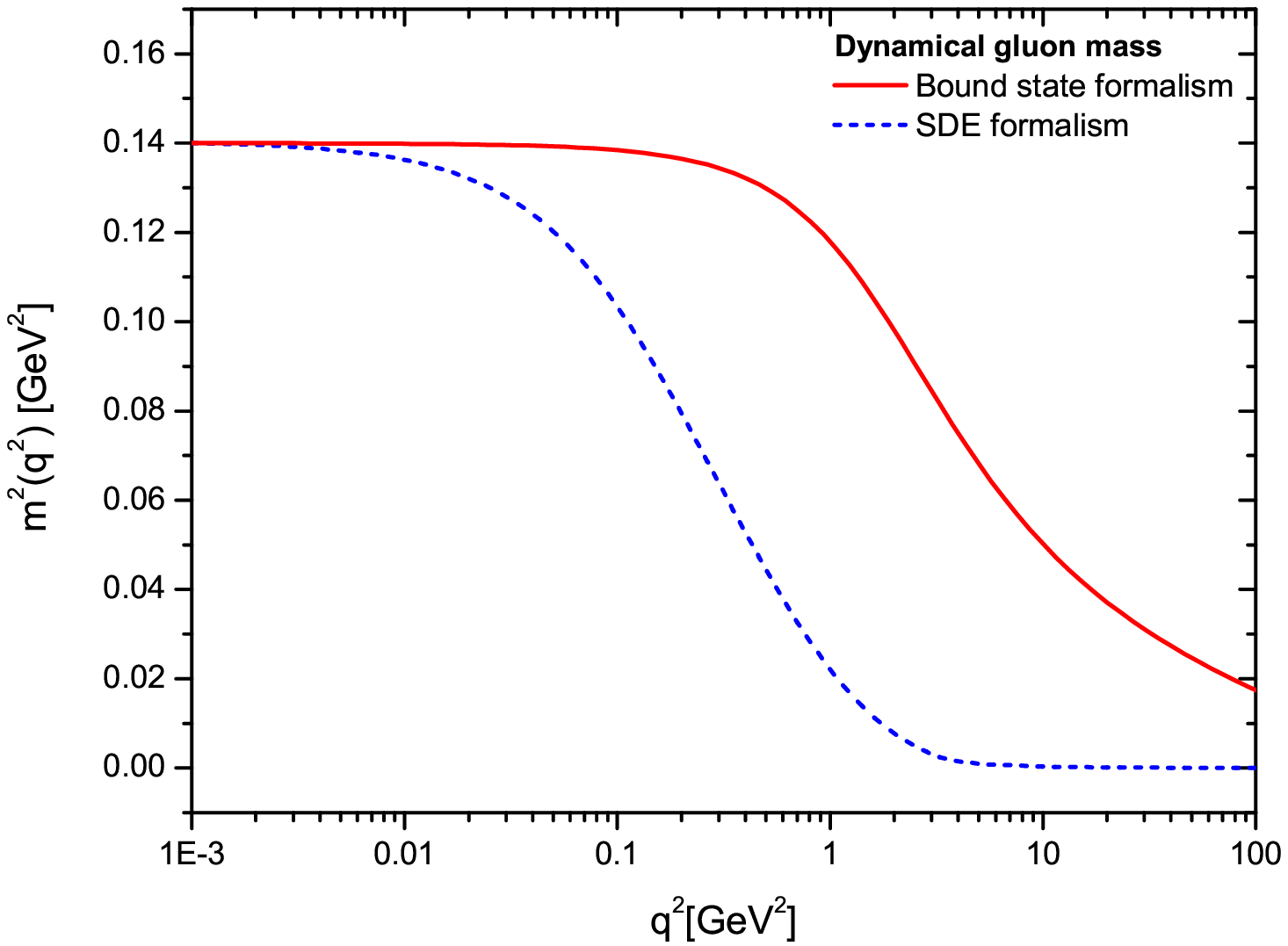}
\end{minipage}
\vspace{-0.25cm}
\caption{\label{solutions} (Left panel): 
The general form of $B'_1(q^2)$ obtained from the numerical solution of \1eq{BSEB1}, in the improved ladder approximation.
(Right panel): The two solutions obtained for the gluon mass within the massless bound-state [red (solid)] and SDE [blue (dotted)] approaches}
\end{figure}

To appreciate this issue in some quantitative detail, recall that, within the massless bound-state formalism, 
the starting point for obtaining the momentum-dependence of the gluon mass 
is \1eq{relB1mass}. Specifically, in the limit $q\rightarrow 0$ one obtains the 
following exact relation for the derivative of the effective gluon mass (Euclidean space)~\cite{Aguilar:2011xe}
\begin{equation}\label{massderivative}
\frac{d m^2(p^2)}{d p^2} = -\widetilde{I}(0)B_1'(p^2),
\end{equation}
where the prime denotes differentiation with respect to $(p+q)^2$ and subsequently taking the 
limit $q\to 0$, \ie
\begin{equation}
B_1'(p^2) \equiv \, \lim_{q\to 0} \left\{ \frac{\partial B_1(q,-p-q,p)}{\partial\, (p+q)^2} \right\} \,.
\label{Der}
\end{equation}
It turns out that 
the function $B_1'(p^2)$ satisfies its own homogeneous BSE [see Section~\ref{mbsf}, Fig.~\ref{BSE} and discussion below], of the general form
\begin{equation}\label{BSEB1}
B_1'(p^2) = \int_k {\cal K}(p,k) B_1'(k^2), 
\end{equation}
where ${\cal K}$ corresponds to the Bethe-Salpeter four-gluon kernel, shown in line $A$ of Fig.~\ref{SDEvsBS} [see also Fig.~\ref{BSE}].
Evidently, the exact treatment of this equation would require the complete knowledge of the four-gluon kernel, 
which, however, is a largely unexplored quantity. 
Therefore, in order to obtain an approximate solution to \1eq{BSEB1}, one resorts to the ``improved''
ladder approximation of this kernel, shown diagrammatically in Fig.~\ref{SDEvsBS}, by dressing the gluon propagators 
but keeping the vertices at tree-level.
Then, the numerical solution of \1eq{BSEB1} yields for $B_1'(q^2)$ the function shown on the left panel of Fig.~\ref{solutions}.

With this information at hand, the  gluon mass may be obtained through direct integration 
of \1eq{massderivative}, namely 
\begin{equation}\label{massderB1}
m^2(q^2) = m^2(0) -\widetilde{I}(0) \int_0^{q^2} dx B_1'(x).
\end{equation}
As for the constant $\widetilde{I}(0)$, using the BQI \1eq{BQItransition}, 
as well as \1eq{funrel} and \1eq{QQmassformula}, one obtains 
\begin{equation}\label{valuezerotran}
\widetilde{I}(0) = F^{-1}(0)I(0) = F^{-1}(0)\sqrt{\frac{m^2(0)}{g^2}} = F^{-1}(0)\sqrt{\frac{\Delta^{-1}(0)}{4\pi\alpha_s}},
\end{equation}
which allow us to estimate $\widetilde{I}(0)$ from the lattice values of the ghost dressing function 
and the gluon propagator at zero momentum, treating $\alpha_s$ as an adjustable parameter. 
Thus, one finally arrives at the dynamical gluon mass shown on the right panel of Fig.~\ref{solutions} 
[red (continuous) curve]~\cite{Aguilar:2011xe}.

Let us now turn to the SDE approach~\cite{Binosi:2012sj}, and compare with the corresponding approximations 
and numerical results. In this case the 
basic dynamical equation is that of \1eq{masseq}, whose central ingredient is the quantity $Y(k^2)$, 
entering into the kernel ${\cal K}_{\s{SD}}$, given in \1eq{SDKernel}. The quantity $Y(k^2)$, 
shown in line $B$ of Fig.~\ref{SDEvsBS}, involves the {\it fully-dressed} three-gluon vertex, which too 
is rather poorly known. As a result, 
in~\cite{Binosi:2012sj}  the quantity $Y(k^2)$ has been approximated by its perturbative (one-loop) 
expression, by replacing the  fully dressed internal gluon  propagators and three-gluon  vertex 
by their tree-level values, as shown in  Fig.~\ref{SDEvsBS}.
Under these approximations, the numerical treatment 
of \1eq{masseq} gives rise to the solution shown on the right panel 
of Fig.~\ref{solutions} [blue (dotted) curve].  

Note that both masses have been normalized in such a way as to coincide at the origin; in particular, 
$m^2(0)= 0.14~{\rm GeV}^2$, which is the saturation point reached by the gluon propagator 
simulated on the lattice, when renormalized at $\mu =4.3~{\rm GeV}$ (the last point available in the ultraviolet tail).  
It is clear from this direct comparison that the two formally equivalent approaches lead to 
qualitatively similar results, which, however, do not coincide, due to the inequivalence of the approximations employed. 

\section{\label{concl} Conclusions}

In this article we have studied in detail the mechanism of gluon mass generation within 
the massless bound-state formalism, which constitutes the formal framework for the  
systematic implementation of the Schwinger mechanism at the level of four-dimensional 
non-Abelian gauge theories.  
The main ingredient of this formalism is the dynamical formation of massless bound-states,
which give rise to effective vertices containing massless poles; these latter vertices  
trigger the Schwinger mechanism, and allow for the gauge-invariant generation of an effective gluon mass.
The principal advantage of this approach is its ability to relate the gluon mass directly 
to quantities that are usually employed in the physics of bound-states, such as transition 
amplitudes and bound-state wave functions, as well as obtaining the dynamical evolution through a BSE
instead of a SDE [see, for example, \1eq{QQmassformula}, \1eq{massderB1} and \1eq{BSEB1}].

A central result of the present work is the formal equivalence 
between the  
massless bound-state formalism and the corresponding approach 
based on the direct study of the SDE of the gluon propagator~\cite{Binosi:2012sj}.
In particular, the powerful relations provided by the STIs of the theory, such as \1eq{relB1mass},
allowed us to demonstrate the exact coincidence of the 
integral equations governing the momentum evolution of the gluon mass in both formalisms.

Note that this formal equivalence, in addition to serving 
as a clear indication of an underlying consistency, opens up the possibility 
of extracting useful, albeit indirect, information on the structure of 
certain quantities appearing in one formalism, 
from results obtained within the other. It seems, for instance, that a nonperturbative approximation for the quantity $Y$ 
(see the SDE of  \1eq{masseq}) is easier to obtain 
than a corresponding approximation for the four-gluon kernel $\cal K$, a central ingredient of the 
BSE in \1eq{BSEB1}. Indeed, $Y$ involves the fully dressed three-gluon vertex, which may be 
partially reconstructed, by means of a gauge-technique Ansatz,
from the STIs that it satisfies \cite{Ball:1980ax}; to be sure, lattice simulation may also be 
valuable in this effort~\cite{Cucchieri:2006tf}.
Instead, nonperturbative approximations for $\cal K$ are much more difficult to obtain with the present technology. 
Therefore, one might be able to use a finer solution for $m^2(q^2)$, obtained from the SDE of \1eq{masseq} 
by going beyond the one-loop approximation for $Y$,  
to infer the behavior of the kernel $\cal K$, at least for some special kinematic configurations.  
This information, in turn, may be helpful in a variety of unrelated studies that involve this particular kernel.  

Even though the derivation of various of the results presented here relies on the 
use of the Landau gauge, it would seem that the general features of the massless bound-state formalism 
remain valid for any value of the gauge-fixing parameter, within the 
general class of linear covariant gauges. It would be therefore most interesting to derive the corresponding 
dynamical equations for different gauges, and explore whether the
gluon mass generation mechanism persists away from the Landau gauge. 
In fact, such studies may be complemented by parallel lattice simulations 
of the corresponding gluon and ghost propagators~\cite{Cucchieri:2011pp,Cucchieri:2011aa}.

In a similar vein, lattice simulations carried out in the BFM~\cite{Binosi:2012st,Cucchieri:2012ii} 
may prove instrumental for verifying explicitly some of the formal relations employed 
throughout this work. For example, the BQI of \1eq{BQIpropagatorsa} may be directly probed, 
for a wide range of momenta, if 
$\widehat{\Delta}(q^2)$ is simulated in the background Landau gauge; for a prediction, see~\cite{Aguilar:2009pp}. 
In fact, even the knowledge of just the corresponding saturation point, $\widehat{\Delta}(0)$,  
would enable one to check the validity of \1eq{BQImasses} at $q^2=0$.

\acknowledgments 
This research is supported by the Spanish MEYC under grant FPA2011-23596.  
We thank A.C.~Aguilar and D.~Binosi for several useful discussions.


\begin{thebibliography}{99}

\bibitem{Cornwall:1981zr}
J.~M.~Cornwall,
Phys.\ Rev.\ D {\bf 26}, 1453 (1982).

\bibitem{Bernard:1982my}
  C.~W.~Bernard,
  Nucl.\ Phys.\ B {\bf 219}, 341 (1983).

\bibitem{Donoghue:1983fy}
  J.~F.~Donoghue,
  Phys.\ Rev.\ D {\bf 29}, 2559 (1984).

\bibitem{Cucchieri:2007md}
A.~Cucchieri and T.~Mendes,
PoS {\bf LAT2007}, 297 (2007).


\bibitem{Cucchieri:2007rg}
A.~Cucchieri and T.~Mendes,
Phys.\ Rev.\ Lett.\  {\bf 100}, 241601 (2008).


\bibitem{Cucchieri:2009zt}
A.~Cucchieri and T.~Mendes,
Phys.\ Rev.\  D {\bf 81}, 016005 (2010).


\bibitem{Bogolubsky:2007ud}
I.~L.~Bogolubsky, E.~M.~Ilgenfritz, M.~Muller-Preussker and A.~Sternbeck,
PoS {LATTICE}, 290 (2007).


\bibitem{Bowman:2007du}
P.~O.~Bowman {\it et al.},
Phys.\ Rev.\  D {\bf 76}, 094505 (2007).


\bibitem{Bogolubsky:2009dc}
I.~L.~Bogolubsky, E.~M.~Ilgenfritz, M.~Muller-Preussker and A.~Sternbeck,
Phys.\ Lett.\  B {\bf 676}, 69 (2009).


\bibitem{Oliveira:2009eh}
O.~Oliveira and P.~J.~Silva,
PoS {\bf LAT2009}, 226 (2009).


\bibitem{Iritani:2009mp} 
  T.~Iritani, H.~Suganuma and H.~Iida,
  Phys.\ Rev.\ D {\bf 80}, 114505 (2009).
  

\bibitem{Aguilar:2008xm} 
  A.~C.~Aguilar, D.~Binosi and J.~Papavassiliou,
  Phys.\ Rev.\ D {\bf 78}, 025010 (2008).


\bibitem{Aguilar:2011ux} 
  A.~C.~Aguilar, D.~Binosi and J.~Papavassiliou,
  Phys.\ Rev.\ D {\bf 84}, 085026 (2011).


\bibitem{Schwinger:1962tn}
  J.~S.~Schwinger,
  Phys.\ Rev.\  {\bf 125}, 397 (1962).


\bibitem{Schwinger:1962tp}
  J.~S.~Schwinger,
  Phys.\ Rev.\  {\bf 128}, 2425 (1962).


\bibitem{Jackiw:1973tr}
  R.~Jackiw and K.~Johnson,
  Phys.\ Rev.\ D {\bf 8}, 2386 (1973).


\bibitem{Jackiw:1973ha}
  R.~Jackiw,
  In *Erice 1973, Proceedings, Laws Of Hadronic Matter*, New York 1975, 225-251 and M I T Cambridge - COO-3069-190 (73,REC.AUG 74) 23p.
  

\bibitem{Cornwall:1973ts}
  J.~M.~Cornwall and R.~E.~Norton,
  Phys.\ Rev.\ D {\bf 8} 3338 (1973).


\bibitem{Eichten:1974et}
E.~Eichten and F.~Feinberg,
Phys.\ Rev.\ D {\bf 10}, 3254 (1974).


\bibitem{Poggio:1974qs}
  E.~C.~Poggio, E.~Tomboulis and S.~H.~Tye,
  Phys.\ Rev.\  D {\bf 11}, 2839 (1975).


\bibitem{Binosi:2012sj} 
  D.~Binosi, D.~Ibanez and J.~Papavassiliou,
  Phys.\ Rev.\ D {\bf 86}, 085033 (2012).


\bibitem{Szczepaniak:2001rg} 
  A.~P.~Szczepaniak and E.~S.~Swanson,
  Phys.\ Rev.\ D {\bf 65}, 025012 (2002).


\bibitem{Szczepaniak:2003ve} 
  A.~P.~Szczepaniak,
  Phys.\ Rev.\ D {\bf 69}, 074031 (2004).


\bibitem{Epple:2007ut} 
  D.~Epple, H.~Reinhardt, W.~Schleifenbaum and A.~P.~Szczepaniak,
  Phys.\ Rev.\ D {\bf 77}, 085007 (2008).


\bibitem{Szczepaniak:2010fe} 
  A.~P.~Szczepaniak and H.~H.~Matevosyan,
  Phys.\ Rev.\ D {\bf 81}, 094007 (2010).


\bibitem{Aguilar:2011xe} 
  A.~C.~Aguilar, D.~Ibanez, V.~Mathieu and J.~Papavassiliou,
  Phys.\ Rev.\ D {\bf 85}, 014018 (2012)


\bibitem{Cornwall:1989gv}
J.~M.~Cornwall and J.~Papavassiliou,
Phys.\ Rev.\  D {\bf 40}, 3474 (1989).


\bibitem{Binosi:2002ft}
D.~Binosi and J.~Papavassiliou,
Phys.\ Rev.\  D {\bf 66}(R), 111901 (2002).


\bibitem{Binosi:2003rr}
D.~Binosi and J.~Papavassiliou,
J.\ Phys.\ G {\bf 30}, 203 (2004).


\bibitem{Binosi:2009qm} 
  D.~Binosi and J.~Papavassiliou,
  Phys.\ Rept.\  {\bf 479}, 1 (2009).


\bibitem{Pilaftsis:1996fh} 
  A.~Pilaftsis,
  Nucl.\ Phys.\ B {\bf 487}, 467 (1997).


\bibitem{Abbott:1980hw}
See, e.g.,  L.~F.~Abbott,
Nucl.\ Phys.\  B {\bf 185}, 189 (1981), and references therein.


\bibitem{Aguilar:2006gr}
A.~C.~Aguilar and J.~Papavassiliou,
JHEP {\bf 0612}, 012 (2006).


\bibitem{Binosi:2007pi}
D.~Binosi and J.~Papavassiliou,
Phys.\ Rev.\  D {\bf 77}(R), 061702 (2008).


\bibitem{Binosi:2008qk}
D.~Binosi and J.~Papavassiliou,
JHEP {\bf 0811}, 063 (2008).


\bibitem{Grassi:1999tp}
P.~A.~Grassi, T.~Hurth and M.~Steinhauser,
Annals Phys.\  {\bf 288}, 197 (2001).


\bibitem{Binosi:2002ez}
D.~Binosi and J.~Papavassiliou,
Phys.\ Rev.\ D {\bf 66}, 025024 (2002).


\bibitem{Batalin:1977pb}
I.~A.~Batalin, G.~A.~Vilkovisky,
Phys.\ Lett.\  {\bf B69}, 309-312 (1977).


\bibitem{Batalin:1981jr}
  I.~A.~Batalin, G.~A.~Vilkovisky,
Phys.\ Lett.\  {\bf B102}, 27-31 (1981).



\bibitem{Aguilar:2002tc} 
  A.~C.~Aguilar, A.~A.~Natale and P.~S.~Rodrigues da Silva,
  Phys.\ Rev.\ Lett.\  {\bf 90}, 152001 (2003).


\bibitem{Aguilar:2004sw} 
  A.~C.~Aguilar and A.~A.~Natale,
  JHEP {\bf 0408}, 057 (2004).


\bibitem{Dudal:2008sp}
  D.~Dudal, J.~A.~Gracey, S.~P.~Sorella, N.~Vandersickel and H.~Verschelde,
  Phys.\ Rev.\ D {\bf 78} (2008) 065047.


\bibitem{Boucaud:2008ky}
  P.~Boucaud, J.~P.~Leroy, A.~Le Yaouanc, J.~Micheli, O.~Pene and J.~Rodriguez-Quintero,
  JHEP {\bf 0806} (2008) 099.


\bibitem{Fischer:2008uz} 
  C.~S.~Fischer, A.~Maas and J.~M.~Pawlowski,
  Annals Phys.\  {\bf 324}, 2408 (2009).


\bibitem{RodriguezQuintero:2010ss}
  J.~Rodriguez-Quintero,
  PoS LC {\bf 2010} (2010) 023.
  

\bibitem{Bicudo:2010wi} 
  P.~Bicudo and O.~Oliveira,
  PoS LATTICE {\bf 2010}, 269 (2010).


\bibitem{Oliveira:2010xc} 
  O.~Oliveira and P.~Bicudo,
  J.\ Phys.\ G G {\bf 38}, 045003 (2011).
  
  
\bibitem{Kondo:2011ab} 
  K.~-I.~Kondo,
  Phys.\ Rev.\ D {\bf 84}, 061702 (2011).
  
  
\bibitem{Qin:2011dd} 
  S.~-x.~Qin, L.~Chang, Y.~-x.~Liu, C.~D.~Roberts and D.~J.~Wilson,
  Phys.\ Rev.\ C {\bf 84}, 042202 (2011).


\bibitem{Gonzalez:2011zc} 
  P.~Gonzalez, V.~Mathieu and V.~Vento,
  Phys.\ Rev.\ D {\bf 84}, 114008 (2011).
  
  
\bibitem{Pennington:2011xs} 
  M.~R.~Pennington and D.~J.~Wilson,
  Phys.\ Rev.\ D {\bf 84}, 119901 (2011).
  

\bibitem{Bashir:2011dp} 
  A.~Bashir, R.~Bermudez, L.~Chang and C.~D.~Roberts,
  Phys.\ Rev.\ C {\bf 85}, 045205 (2012).


\bibitem{Bashir:2012fs} 
  A.~Bashir, L.~Chang, I.~C.~Cloet, B.~El-Bennich, Y.~-X.~Liu, C.~D.~Roberts and P.~C.~Tandy,
  Commun.\ Theor.\ Phys.\  {\bf 58}, 79 (2012).


\bibitem{Kondo:2012ri} 
  K.~-I.~Kondo, K.~Suzuki, H.~Fukamachi, S.~Nishino and T.~Shinohara,
  arXiv:1209.3994 [hep-th].


\bibitem{Strauss:2012dg} 
  S.~Strauss, C.~S.~Fischer and C.~Kellermann,
  arXiv:1208.6239 [hep-ph].

\bibitem{Ball:1980ax}
J.~S.~Ball and T.~W.~Chiu,
Phys.\ Rev.\  D {\bf 22}, 2550 (1980)
[Erratum-ibid.\  D {\bf 23}, 3085 (1981)].


\bibitem{Aguilar:2009nf}
  A.~C.~Aguilar, D.~Binosi, J.~Papavassiliou and J.~Rodriguez-Quintero,
  Phys.\ Rev.\ D {\bf 80} (2009) 085018.


\bibitem{LlewellynSmith:1969az}
  C.~H.~Llewellyn-Smith,
  Annals Phys.\  {\bf 53} (1969) 521.
  

\bibitem{Maris:2003vk}
  P.~Maris and C.~D.~Roberts,
  Int.\ J.\ Mod.\ Phys.\ E {\bf 12} (2003) 297.


\bibitem{Bjorken:1979dk}
  J.~D.~Bjorken and S.~D.~Drell, ``Relativistic Quantum Field Theory'', McGraw-Hill Inc (1965), 
chapter 19. 


\bibitem{Pascual:1984zb} 
  P.~Pascual and R.~Tarrach,
  ``Qcd: Renormalization For The Practitioner,''
  Lect.\ Notes Phys.\  {\bf 194}, 1 (1984).


\bibitem{Davydychev:1996pb} 
  A.~I.~Davydychev, P.~Osland and O.~V.~Tarasov,
  Phys.\ Rev.\ D {\bf 54}, 4087 (1996)
  [Erratum-ibid.\ D {\bf 59}, 109901 (1999)].


\bibitem{Grassi:2004yq}
  P.~A.~Grassi, T.~Hurth and A.~Quadri,
  Phys.\ Rev.\ D {\bf 70} (2004) 105014.


\bibitem{Aguilar:2009pp}
  A.~C.~Aguilar, D.~Binosi and J.~Papavassiliou,
  JHEP {\bf 0911} (2009) 066.


\bibitem{Aguilar:2010gm}
  A.~C.~Aguilar, D.~Binosi and J.~Papavassiliou,
  JHEP {\bf 1007} (2010) 002.


\bibitem{Ibanez:2011hc} 
  D.~Ibanez,
  PoS QCD {\bf -TNT-II}, 025 (2011).


\bibitem{Binosi:2011wi}
  D.~Binosi and J.~Papavassiliou,
  JHEP {\bf 1103} (2011) 121.


\bibitem{Cornwall:1985bg} 
  J.~M.~Cornwall and W.~-S.~Hou,
  Phys.\ Rev.\ D {\bf 34}, 585 (1986).


\bibitem{Cucchieri:2006tf} 
  A.~Cucchieri, A.~Maas and T.~Mendes,
  Phys.\ Rev.\ D {\bf 74}, 014503 (2006).


\bibitem{Cucchieri:2011pp} 
  A.~Cucchieri, T.~Mendes, G.~M.~Nakamura and E.~M.~S.~Santos,
  PoS FACESQCD {\bf }, 026 (2010).


\bibitem{Cucchieri:2011aa} 
  A.~Cucchieri, T.~Mendes, G.~M.~Nakamura and E.~M.~S.~Santos,
  AIP Conf.\ Proc.\  {\bf 1354}, 45 (2011).


\bibitem{Binosi:2012st} 
  D.~Binosi and A.~Quadri,
  Phys.\ Rev.\ D {\bf 85}, 121702 (2012).


\bibitem{Cucchieri:2012ii} 
  A.~Cucchieri and T.~Mendes,
  Phys.\ Rev.\ D {\bf 86}, 071503 (2012).


\end{thebibliography}
\end{document}